\documentclass[oneside,a4paper,english]{article}
\usepackage[a4paper,left=1in,right=1in,top=1in,bottom=1in]{geometry}
\usepackage{graphicx}
\usepackage{epsfig}
\usepackage{color}
\usepackage{amsmath,amsfonts}
\usepackage{subfigure}
\usepackage{multirow}
\usepackage{algorithm2e}

\usepackage{booktabs}

\LinesNumbered

\graphicspath{{figs/}}

\title{A stable numerical strategy for Reynolds-Rayleigh-Plesset coupling}

\author{Alfredo Jaramillo$^a$, Gustavo C. Buscaglia$^a$\\
	{\small
		(a) Instituto de Ci\^encias Matem\'aticas e de Computa\c{c}\~ao,
		Universidade de S\~ao Paulo, 13560-970 S\~ao Carlos, Brazil
}}

\newcommand{\parder}[2]{\frac{\partial #1}{\partial #2}}
\newcommand{\pcav}{p_{\textnormal{\scriptsize{cav}}}}
\newcommand{\pref}{p_{\textnormal{\scriptsize{ref}}}}
\newcommand{\Rref}{R_{\textnormal{\scriptsize{ref}}}}
\newcommand{\fref}{f_{\textnormal{\scriptsize{ref}}}}

\begin{document}
	\vspace{3cm}
	
	\maketitle
	
	%% To set PDF metadata: uncomment and replace fields in
	%% UPPERCASE with appropriate values. 
	%% \pdfinfo{ /Title (ARTICLE TITLE)
	%%           /Author            (AUTHORS)
	%%           /Subject           (CONGRESS-TRACK)
	%%           /Keywords          (KEYWORDS) }
	%%
	%% For instance
	%% \pdfinfo{ /Title (A new model for multi-phase flow)
	%%           /Author            (Spone B. and Star P.)
	%%           /Subject           (Fluid Mechanics)
	%%           /Keywords          (multiphase flow, air-liquid mixtures) }
	
	\begin{abstract}
		The coupling of Reynolds and Rayleigh-Plesset equations has been used in several works to simulate lubricated devices considering cavitation. The numerical strategies proposed so far are variants of a staggered strategy where Reynolds equation is solved considering the bubble dynamics frozen, and then the Rayleigh-Plesset equation is solved to update the bubble radius with the pressure frozen. We show that this strategy has severe stability issues and a stable methodology is proposed. The proposed methodology performance is assessed on two physical settings. The first one concerns the propagation of a decompression wave along a fracture considering the presence of cavitation nuclei. The second one is a typical journal bearing, in which the coupled model is compared with the Elrod-Adams model.
		
		\medskip
		
		%{\bf Keywords:} Piston ring, cylinder liner, hydrodynamic lubrication,
		%cavitation, mass-conservative model, mesh convergence,
		%textured liner, pocketed surface, friction reduction, ring dynamics.
		
		
	\end{abstract}
	
	\section{Introduction}\label{sec:introduction}
	
	Cavitation modeling is a challenging issue when studying the hydrodynamics 
	of lubricated devices \cite{dowson1979,braun2010}. It is experimentally
	known that gases (small or large bubbles of air or vapor) appear in the
	liquid lubricant in regions where otherwise the pressure would be
	negative. The volume occupied by these gas bubbles affects
	the pressure field, to the point of preventing it from
	developing negative regions.
	It is customary to think of the whole fluid
	(lubricant + gas) as a mixture for which it is possible to define
	effective fields of pressure ($p$), density ($\rho$) and viscosity ($\mu$).
	These three fields are linked by the well-known Reynolds equation,
	which expresses
	the conservation of mass and must thus hold for the cavitated mixture as
	well as for the pure lubricant.
	
	Notice, however, that while in problems in which the lubricant is
	free of gases the density and viscosity are given material data, in problems
	with significant gas content $\rho$ and $\mu$ are two additional
	unknown fields (totalling three with $p$). The overall behavior of
	the mixture exhibits low-density regions (i.e., regions where the
	fraction of gas is high) appearing at places where otherwise the
	pressure would be negative, in such a way that the overall pressure
	field does not exhibit negative (or very negative) values.
	
	These low-density regions
	are usually called {\em cavitated} regions, though the gas may have
	appeared there by different mechanisms: cavitation itself (the growth
	of bubbles of vapor), growth of bubbles of dissolved gases, ingestion
	of air from the atmosphere surrounding the lubricated device, etc.
	
	Many mathematical models have been developed over the years to predict
	the behavior of lubricated devices that exhibit cavitation, and most
	of them have been implemented numerically (see, e.g., \cite{braun2010}).
	The most widely used models assume that
	the data (geometry, fluid properties) and the resulting flow are
	smooth in time, with time scales governed by the macroscopic
	dynamics of the device.
	In particular, the fast transients inherent to the dynamics
	of microscopic bubbles, though being the physical origin of cavitation,
	are averaged out of the model.
	To accomplish this, these models propose phenomenological
	laws relating $\rho$, $p$ and $\mu$. These laws vary from very
	simple to highly sophisticated and nonlocal, and
	may involve one or more additional (e.g., auxiliary) variables.
	
	A representative example of the aforementioned models is
	Vijayaraghavan and Keith's bulk \cite{Vijayaraghavan1989,Vaidyanathan1989} compressibility modulus model.
	Without going into the details, it essentially postulates that
	\begin{equation}
	p = \begin{cases}
	p_{\mbox{\scriptsize{cav}}} + \beta\,\ln \left ( \frac{\rho}{\rho_\ell} \right )
	& \mbox{if}~p\geq p_{\mbox{\scriptsize{cav}}} \mbox{ and }
	\rho \geq \rho_\ell \\
	p_{\mbox{\scriptsize{cav}}} & \mbox{otherwise}
	\end{cases}\quad ,
	\label{eq:p_vk}
	\end{equation}
	where $\rho_\ell$ is the liquid density and $p_{\mbox{\scriptsize{cav}}}$ and
	$\beta$ are constants. Another example is the Elrod-Adams model, which
	here is considered in the mathematical form made precise by Bayada and coworkers \cite{Bayada1982}
	and which can be viewed, to some extent, as a limit of (\ref{eq:p_vk})
	for $\beta \to +\infty$.
	
	In recent years, detailed measurement and simulation of lubricated devices with wide ranges in their spatial and temporal scales has become affordable \cite{checo2016,checo2014a,Checo2017}. Micrometric features of the lubricated surfaces can now be incorporated into the simulated geometry, down to the roughness scale. These micrometric spatial features of the two lubricated surfaces, being in sliding relative motion, generate rapid transients in the flow. This reason, among others, has lately revived the interest on models that take the microscopic dynamics of the incipient cavitation bubbles (or {\em nuclei}) into account. We refer to them as {\em bubble-dynamics-based} models, and they are the focus of this contribution.
	
	To our knowledge, this kind of models was first used by T\o{}nder \cite{Tonder1977} for Michell Bearings. After that article, and progressively augmenting the complexity of the gas-nuclei dynamics, several works have been published concerning tilting-pad thrust bearings \cite{Smith1980}, journal bearings \cite{Natsumeda1987,Someya2003,Snyder2016,Snyder2017}, squeeze film dampers \cite{Gehannin2009,Gehannin2016} and parallel plates \cite{Geike2009a,Geike2009b}.
	
	In bubble-dynamics-based models the gas-nuclei dynamics intervenes in the
	Reynolds equation through the fraction variable
	$$
	\alpha=\frac{\mbox{volume of gas}}{\mbox{volume of gas and liquid}}~,
	$$
	from which
	the  density $\rho(\alpha)$ and viscosity $\mu(\alpha)$ of the mixture
	are obtained. The gas fraction $\alpha$, in turn, depends on the size and number of nuclei, which are modeled as spherical bubbles of radius $R$
	and assumed to obey the well-known Rayleigh-Plesset equation of bubble dynamics.
	
	The resulting mathematical equations exhibit the so-called {\em Reynolds-Rayleigh-Plesset} (RRP) coupling. By this it is meant that the coefficients of the
	equation that determines $p$ (i.e., the Reynolds equation) depend on the 
	local values of $R$, while the driving force of the equation that governs
	the dynamics of $R$ (i.e., the Rayleigh-Plesset equation) depends on $p$.
	
	In this work a stable numerical approach for RRP coupling is presented, designed for problems in which inertia can be neglected. It can be seen as an extension of the work by Geike and Popov \cite{Geike2009a,Geike2009b}, who only considered the very specific geometry of parallel plates.
	
	In Section \ref{sec:rp-equation} the Rayleigh-Plesset equation is presented along some of its properties, and the simplified version which is used in this work. Section \ref{sec:coupling} is dedicated to the definitions needed to couple Reynolds and Rayleigh-Plesset equations, and to present the two numerical methods compared in this work. Numerical results are presented in Section \ref{sec:numerical_results}, first for an original problem where the pressure build-up is generated only by expansion/compression of the bubbles both in one-dimensional (1D) and two-dimensional (2D) settings. It is shown that the proposed methodology allows to perform simulations and discover features of the Reynold-Rayleigh-Plesset coupling that are not possible to be computed with the method found in the literature. After that, numerical results regarding the Journal Bearing are presented showing the robustness of the proposed method when varying operational conditions and fluid properties. Concluding remarks are given in the last section.

	\section{The field Rayleigh-Plesset equation}
	\label{sec:rp-equation}
\begin{figure}[hb]
		\centering 
		\def\svgwidth{\textwidth}	
		\includegraphics[scale=0.7]{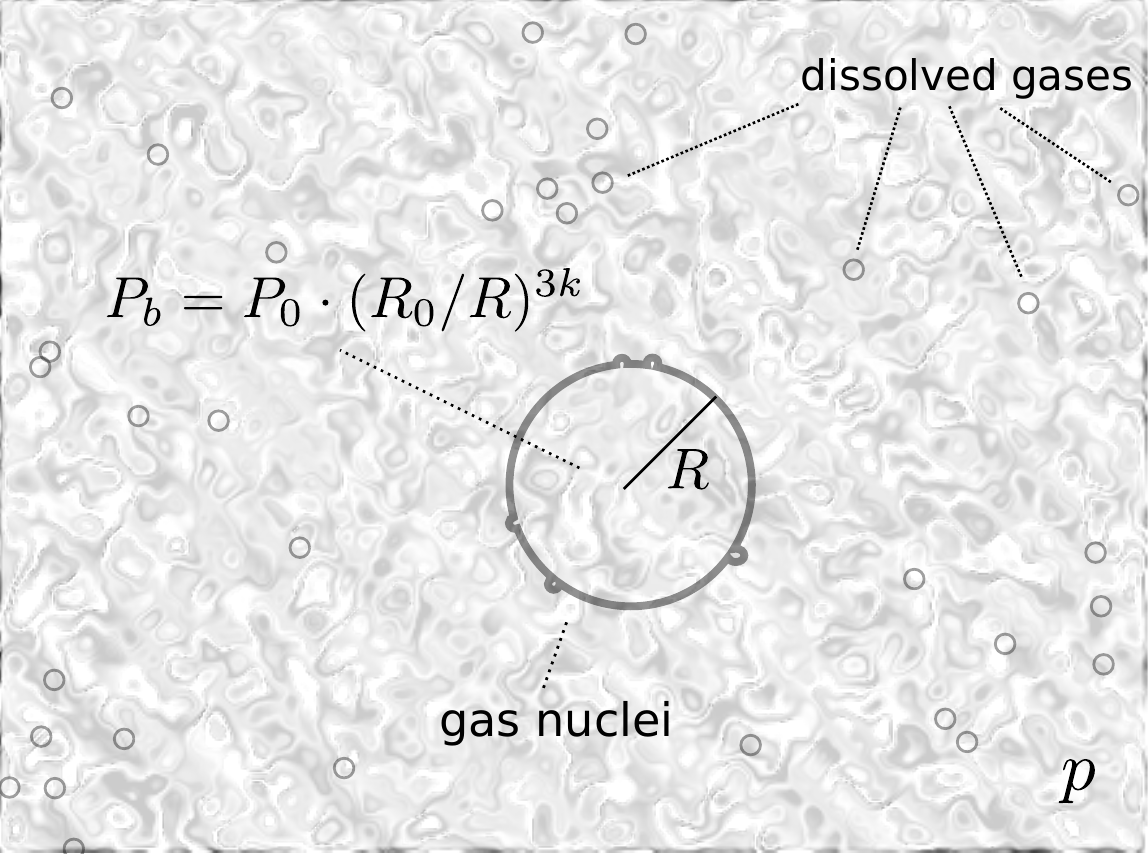}
		\caption[Illustration of gaseous cavitation]{Illustration of an idealized bubble of radius $R$ immersed in a Newtonian fluid. The formula for the inner pressure of the bubble, $P_b$, assumes adiabatic behavior. The pressure far away from the bubble is denoted by $p$.}\label{fig:gaseous_cavitation}
	\end{figure}

	The evolution of a small spherical gas bubble immersed in a Newtonian fluid (as illustrated in Fig. \ref{fig:gaseous_cavitation}) in adiabatic conditions is governed by the Rayleigh-Plesset (RP) equation \cite{Brennen1995}, which reads 
	\begin{equation}
	\rho_\ell\left(\frac{3}{2}\dot{R}^2+R\ddot{R}\right)=-\left(\mu_\ell+\frac{\kappa^s}{R}\right)\frac{4\dot{R}}{R}+F(R)-p\label{eq:rayleigh-plesset-full}
	\end{equation}
	where $\dot{R}=\frac{dR}{dt}$ is the total time derivative (following the bubble), $R(t)$ is the radius, $\rho_\ell$, $\mu_\ell$ correspond to the fluid density and viscosity  respectively, $\kappa^s$ is the surface dilatational viscosity \cite{Snyder2016}, $p$ is the liquid pressure far away from the bubble and $F(R)$ reads
	\begin{equation}
	F(R)=P_0\left(\frac{R_0}{R}\right)^{3k}-\frac{2\sigma}{R}~,\label{eq:young-laplace}
	\end{equation}
	where $P_0$ is the inner pressure of the bubble when its radius is equal to $R_0$ and $\sigma$ is the surface tension. In the right-hand side of the last equation the first term models the pressure of the gas contained in the bubble (in this work the polytropic exponent is fixed to $k=1.4$) and the latter term corresponds to the surface pressure jump. 
	
\begin{figure}[htb]
		\centering 
		\def\svgwidth{\textwidth}	
		\includegraphics[scale=0.95]{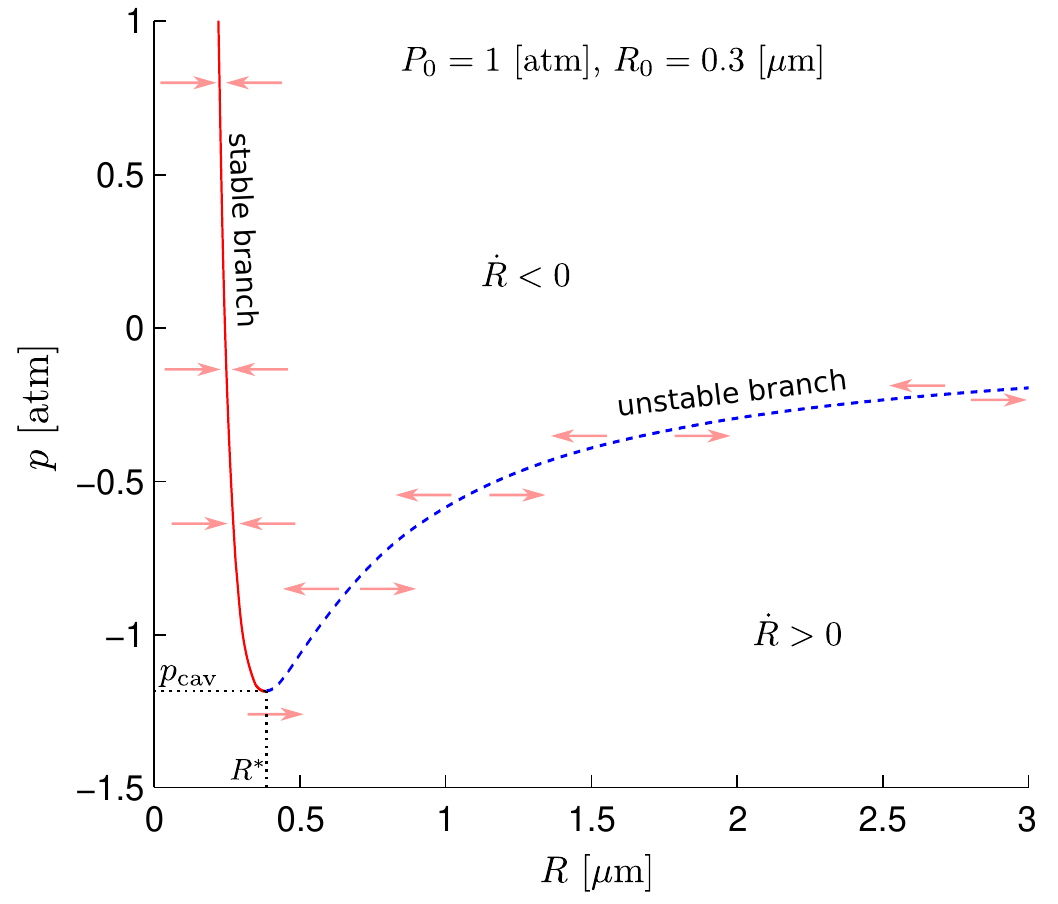}
		\caption{The function $F(R)$ for a typical pair  $(P_0,R_0)$. The red continuous line and the dashed blue line correspond to the stable and unstable branch respectively. The arrows indicate the sign of $\dot{R}$ at each region separated by $F(R)$. Also $R^*=\underset{R>0}{\arg\min}~ F(R)$ and $\pcav=F(R^*)$.}\label{fig:sec:qualitative:rayleigh-plesset-p-vs-R}
	\end{figure}

	Hereafter the inertial terms in Eq. \eqref{eq:rayleigh-plesset-full} are assumed to be negligible (e.g., \cite{Snyder2016}), which leads to the inertialess Rayleigh-Plesset equation 
	\begin{equation}
	\frac{dR}{dt}=G(R)\left(F(R)-p\right),\label{eq:rayleigh-plesset-withouthinertia}
	\end{equation}
	with
	\begin{equation*}
	G(R)=	\frac{R}{4\mu_\ell+4\kappa^s/R}.\label{eq:def-G}
	\end{equation*}
	In Fig. \ref{fig:sec:qualitative:rayleigh-plesset-p-vs-R} the typical shape of the function $F$ (e.g., \cite{Brennen1995}) is shown. Considering some value of $p$ constant in time, from Eq. \eqref{eq:rayleigh-plesset-withouthinertia} is can be noticed that for a bubble to be in equilibrium ($\dot{R}=0$) at such $p_e$ there must exist some radius $R_e=R_e(p_e)$ such that $F(R_e)=p_e$, i.e.,
	\begin{equation}
	p_e=P_0\left(\frac{R_0}{R_e}\right)^{3k}-\frac{2\sigma}{R_e}.\label{eq:equilibrium}
	\end{equation}
	This is possible only if $p_e$ is higher that the minimum value of $F$. Thus, we denote
	\begin{equation}
	R^*=\underset{R>0}{\arg\min}~ F(R)=\left[\frac{3kP_0R_0^{3k}}{2\sigma}\right]^{\frac{1}{3k-1}},\label{eq:Rc}
	\end{equation}
	and
	\begin{equation}
	\qquad \pcav=F(R^*).\label{eq:pcav}
	\end{equation}
	From Eq. \eqref{eq:rayleigh-plesset-withouthinertia} observe for $R_e<R^*$ the equilibrium states are stable while for $R_e>R^*$ the equilibrium states are unstable. Moreover, if $p$ is below $\pcav$ the bubbles grow
	monotonically, only stopping if the pressure is increased. Hereafter, we name $\pcav$ as \emph{cavitation pressure}.
	
	The single-bubble equation (\ref{eq:rayleigh-plesset-withouthinertia}) is transformed into a field equation by assuming that there is a large number of bubbles, and adopting spatial, temporal or probabilistic averaging \cite{Drew1999}. At this point the complexity can grow substantially if there exist bubbles of different sizes at any fluid point and instant, in which case one must adopt a {\em polydisperse} model \cite{Drew1999}.
	Such models are based on a population balance equation for the
	so-called {\em bubble density function} $f_b(r,x,t)$, defined such that the number of bubbles
	per unit volume at $x$ and $t$ with radius between $r$ and $r+dr$ is $f_b(r,x,t)dr$
	\cite{gccq94,cc13}. Techniques for numerically handling polydisperse models can
	be found in the literature \cite{cdbl99,Ramkrishna00,cc13,clc16}.
	
	In this contribution, as has been the rule in previous works on RRP coupling,
	we assume the flow to be {\em monodisperse}. That is, we assume that in the
	vicinity of any point $x$, at each time $t$, there exist bubbles of one and
	only one radius, $R(x,t)$. In terms of the bubble density function,
	\begin{equation}
	f_b(r,x,t)=n_b(x,t)\,\delta(r-R(x,t))~,
	\end{equation}
	where $n_b$ is the number concentration of bubbles. In monodisperse models,
	the field unknowns are $R$ and $n_b$. An equation for $R$ is readily obtained
	from the Rayleigh-Plesset equation. Denoting by $\vec{V}=(u,v,w)$ the velocity
	field of the bubbles and now considering the {\em field} $R(x,t)$ one has, from
	(\ref{eq:rayleigh-plesset-withouthinertia}),
	\begin{equation}
	{\parder{R}{t}}+\vec{V}\cdot \nabla R= G(R)\left(F(R)-p\right)~.
	\label{eqfieldr}
	\end{equation}
	The equation for $n_b$, on the other hand, assuming there is no coalescence
	or rupture of bubbles, simply reads
	\begin{equation}
	\parder{n_b}{t}+\nabla \cdot \left ( \vec{V}\,n_b \right ) = 0~.
	\label{eqnb}
	\end{equation}
	Unlike (\ref{eqfieldr}), the equation above
	does not involve the pressure field, so that
	if the bubble velocity $\vec{V}$ is known (\ref{eqnb}) can be solved
	separately and $n_b(x,t)$ considered a given datum. Also, in some
	cases algebraic expressions for $n_b$ can be built, as is shown
	in one of the numerical examples. For these reasons, we consider hereafter
	that $n_b$ is given, so that the {\em field Rayleigh-Plesset} equation
	(\ref{eqfieldr}) is the only equation we are
	left with. Because it is a transport equation, it requires an initial
	condition $R(x,t=0)$ and a boundary condition at {\em inflow} boundaries
	of $\vec{V}$.
	
	Notice that (\ref{eqfieldr}) involves
	two unknowns ($R$ and $p$), so that an additional equation is needed
	to close the system. This equation is the {\em compressible Reynolds
		equation} and will be introduced in the next section.
	
	For later use, let us point out that assuming the bubbles to
	be spherical gives the algebraic relation (e.g., \cite{Singhal2002,Zwart2004})
	\begin{equation}
	\alpha =  \frac{4\pi n_b}{3} {R}^3~
	\label{eq:alpha2}
	\end{equation}
	which links the gas volume fraction $\alpha$ to the main unknown $R$.
	
	\bigskip
	
	\noindent{\em Remark:} Because the liquid is incompressible, it can
	be argued that the number of bubbles per unit volume {\em of liquid},
	that we denote here by $n_{b\ell}$, remains constant. 
	This is certainly valid if the bubbles are very
	small and they move with the local velocity of the liquid. 
	Under this assumption equation (\ref{eqnb}) can be replaced by
	$n_b/(1-\alpha)=n_{b\ell}=\mbox{ constant}$, which leads to
	\begin{equation}
	n_b=\frac{n_{b\ell}}{1+\frac{4\pi R^3}{3} n_{b\ell}}
	\label{eqnbl1}
	\end{equation}
	and thus (e.g., \cite{Schnerr2001}) to
	\begin{equation}
	\alpha = \frac{\frac{4\pi R^3}{3} n_{b\ell}}{1+\frac{4\pi R^3}{3} n_{b\ell}}~.
	\label{eqnbl2}
	\end{equation}
	The constant $n_{b\ell}$ represents the number of microbubble nuclei
	present in the liquid lubricant.
	%\end{document}
	
	\section{Coupling Reynolds and Rayleigh-Plesset equations} \label{sec:coupling}
	
	We consider two surfaces in close proximity that are in relative motion at speed $U$ along the $x_1$-axis. The gap between these surfaces, represented by the function $h(x,t)$, is assumed to be filled by a Newtonian fluid (which can be a mixture) of density $\rho$ and viscosity $\mu$. The functions $h(x,t)$ and $D_t h(x,t)=\parder{h}{t}(x,t)$ are also assumed to be known. To solve for the hydrodynamical pressure $p$ we use the compressible Reynolds equation,
	\begin{equation}
	\nabla\cdot\left(\frac{\rho h^3}{12\mu }\nabla p \right)=\frac{U}{2}\parder{\rho h}{x_1}+\rho\parder{ h}{t}+h\parder{\rho }{t}~,\label{eq:reynolds}
	\end{equation}
	along with the boundary conditions
	\begin{equation}
	p=p_\partial,~\mbox{on}\,\partial \Omega_D,\qquad \nabla p\cdot \hat{n}=g,~\mbox{on}\,\partial \Omega_N~,
	\label{eq:rbc}
	\end{equation}
	where $\hat{n}$ is the unitary vector pointing outwards of $\Omega$ at each point of its boundary  $\partial \Omega=\partial \Omega_D\cup \partial \Omega_N$, and $u$, $g$ are smooth given functions. 
	
	The mixture density and viscosity are assumed to depend on $\alpha$ (and thus on $R$, from (\ref{eq:alpha2})) according to
	\begin{equation}
	\rho(\alpha) = (1-\alpha)\rho_\ell + \alpha \rho_g~,
	\label{eq:rho_alpha}
	\end{equation}
	\begin{equation}
	\mu(\alpha) = (1-\alpha)\mu_\ell + \alpha \mu_g~.
	\label{eq:mu_alpha}
	\end{equation}
	
	The {\em coupled RRP problem} thus consists of determining the fields
	$R(x,t)$ and $p(x,t)$ such that (\ref{eqfieldr})
	and (\ref{eq:reynolds}) are simultaneously satisfied for all $x$ in the
	domain and all $t>0$. The boundary conditions are (\ref{eq:rbc}) and
	the value of $R$ at inflow boundaries. An initial condition for
	$R$ is also enforced.
	
	Notice that the coefficients $\rho$ and $\mu$
	depend on the local value of $R$ through
	equations (\ref{eq:rho_alpha}), (\ref{eq:mu_alpha})
	and (\ref{eq:alpha2}). The two equations are thus coupled and none of
	them can be solved independently of the other.
	
	\subsection{Discretization: The Staggered scheme}
	Assuming $\Omega$ to be a rectangular domain, Eq. (\ref{eq:reynolds}) is discretized by means of a Finite Volume scheme using rectangular cells of length $\Delta x_1$ ($\Delta x_2$) along the $x_1$-axis ($x_2$-axis). The coordinates $(x_i,y_j)$ correspond to the cells' centers. Using a constant time step $\Delta t$, we denote $t^n=n\,\Delta t$ for $0=1\ldots N$.
	
	Consider that $R_{ij}^n$ and $R_{ij}^{n-1}$, for all cells,
	have already been calculated. The {\em Staggered scheme}  (presented, e.g., in
	\cite{Snyder2016,meng2016}) of discretization of the coupled RRP problem 
	is defined by the following equations:
	
	\bigskip
	
	{\noindent{\bf Stage 1: Computation of $p^n$}}
	\begin{multline}
	\frac{c_{i-\frac{1}{2},j}\,p^n_{i-1,j}-\left(c_{i-\frac{1}{2},j}+c_{i+\frac{1}{2},j}\right)p^n_{i,j}+c_{i+\frac{1}{2},j}\,p^n_{i+1,j}}{\Delta x_1^2} +\\ 
	+\frac{c_{i,j-\frac{1}{2}}\,p^n_{i,j-1}-\left(c_{i,j-\frac{1}{2}}+c_{i,j+\frac{1}{2}}\right)p^n_{i,j}+c_{i,j+\frac{1}{2}}\,p^n_{i,j+1}}{\Delta x_2^2} =\\= \frac{U}{2}\left(\frac{\rho^n_{i,j}h^n_{i,j}-\rho^n_{i-1,j}h^n_{i-1,j}}{\Delta x_1}\right)+\rho^n_{i,j}\left(D_th\right)^n_{i,j}+ h^n_{i,j}\,\frac{\rho_{i,j}^n-\rho_{i,j}^{n-1}}{\Delta t}
	\label{eq:reynolds_discrete}
	\end{multline}
	with $$c_{i\pm \frac{1}{2},j}=\frac{\rho^n_{i,j}\left(h_{i,j}^n\right)^3/(12\mu)+\rho^n_{i\pm 1,j}\left(h_{i\pm 1,j}^n\right)^3/(12\mu)}{2},$$
	$$c_{i,j\pm \frac{1}{2}}=\frac{\rho^n_{i,j}\left(h_{i,j}^n\right)^3/(12\mu)+\rho^n_{i,j\pm 1}\left(h_{i,j\pm 1}^n\right)^3/(12\mu)}{2}.$$
	Eq. (\ref{eq:reynolds_discrete}) is used at each cell that does not belong to the domain, while for the cells belonging to the boundary of $\Omega$ we have
	\begin{equation}
	p_{i,j}^n=\left.p_\partial\right|_{x_{i},y_j}~ \mbox{for }(x_i,y_j)\in \partial \Omega_D\label{eq:discrete-boundary-condition_D}
	\end{equation}
	and
	\begin{equation}
	\left(D_{\hat{n}}\, p\right)_{i,j}^n=0~ \mbox{for }(x_i,y_j)\in \partial \Omega_N,\label{eq:discrete-boundary-condition_N}
	\end{equation}
	where $\left(D_{\hat{n}}\, p\right)_{i,j}^n$ is a first order approximation of the normal derivative at the boundary $\partial \Omega_N$.
	
	\bigskip
	
	\noindent{\bf Stage 2: Computation of $R^{n+1}$}
	
	In the lubrication approximation $w$ is neglected in (\ref{eqfieldr}), turning
	it into a two-dimensional transport equation. Among many possibilities, our
	implementation adopted the following implicit scheme:
	\begin{equation}
	R_{i,j}^{n+1}=R_{i,j}^{n}+\Delta t \left\{G\left(R_{i,j}^{n+1}\right)\left(F\left(R_{i,j}^{n+1}\right)-p^{n}_{i,j}\right)-\left(u\,D_1R\right)^{n+1}_{i,j}-\left(v\,D_2R\right)^{n+1}_{i,j}\right\}\label{eq:rayleigh-plesset-discrete},
	\end{equation}
	where $D_1=\parder{}{x_1}$, $D_2=\parder{}{x_2}$, and the convective terms $u\,D_1R$ and $v\,D_2R$ are discretized by means of an upwind scheme. Notice that
	(\ref{eq:rayleigh-plesset-discrete}) is a discretization of (\ref{eqfieldr}),
	which is a transport equation but {\em not} a conservation law because
	$\nabla \cdot \vec{V}\neq 0$ in general. Gehannin et al \cite{Gehannin2016}
	discuss its treatment by the finite volume method in the context of RRP coupling.
	
	After the two stages above all variables have been updated and
	the code may proceed to the next time step. The scheme is called staggered
	because the {\em Reynolds solver} (Stage 1) computes $p^n$ assuming
	that $R=R^n$ and $(\partial \rho/\partial t)^n=(\rho^n-\rho^{n-1})/\Delta t$,
	both already calculated, and then the {\em Rayleigh-Plesset solver} (Stage 2)
	computes $R^{n+1}$ with the pressure fixed at $p=p^n$. The implementation
	of the staggered scheme of RRP coupling is straightforward, since no
	modification is required in the Reynolds solver. It should be noticed,
	however, that $\partial \rho/\partial t$ is approximated with information
	from the {\em previous} time step. Both $R^{n-1}$ and $R^n$ are required
	to compute $R^{n+1}$. This gives rise to initialization issues and also,
	as shown along the next sections, to stability limits. A method that
	only requires information of the current time step is described next.
	
	\subsection{Discretization: The Single-step scheme}
	This is the method proposed in this work, which can be seen as an adaptation of that in \cite{Geike2009a}. The basic idea is to compute $(\partial \rho/\partial t)^n$ using only information about $R^n$ (and thus $\alpha^n$ and $\rho^n$). From the chain rule, assuming
	$\rho_\ell$ constant, we have
	\begin{equation}
	\frac{\partial\rho}{\partial t}=
	-(\rho_\ell-\rho_g)\frac{\partial \alpha}{\partial t} + \alpha\,\frac{\partial \rho_g}{\partial t}~=~-4\pi(\rho_\ell-\rho_g)n_b\,R^2\, \frac{\partial R}{\partial t}+\frac{4\pi(\rho_\ell-\rho_g)R^3}{3}\,
	\frac{\partial n_b}{\partial t}
	+ \alpha\,\frac{\partial \rho_g}{\partial t} ~.
	\end{equation}
	To simplify the exposition, as done by other authors \cite{Natsumeda1987, Gehannin2009,Geike2009a,Geike2009b},
	one may assume that
	$\partial \rho_g/\partial t = 0$ and that $\partial n_b/\partial t=0$ \cite{Natsumeda1987,Snyder2016,Gehannin2009,Geike2009a,Geike2009b}.
	Denoting $K(R)=4\pi (\rho_\ell-\rho_g) n_b R^2$, one arrives at
	\begin{equation}
	\frac{\partial\rho}{\partial t}~=~-K(R)\, \frac{\partial R}{\partial t}~.
	\label{eqdrhodt}
	\end{equation}
	It is worth mentioning that $\partial \rho_g/\partial t$
	is in fact equal to $-(3\rho_0R_0^3/R^4)\,\partial R/\partial t$, where
	$\rho_0$ is the density when $R=R_0$. Incorporating this effect in the
	model amounts to adding the term $3\alpha\rho_0R_0^3/R^4$ to the definition of
	$K(R)$. Similarly, if one adopts the model (\ref{eqnbl1}) the factor $K(R)$
	changes to $\overline{K}(R)=(1+\alpha)\,K(R)$. These changes in $K(R)$ have
	no numerical consequences, so that the algorithm presented below can
	be applied in any case.
	
	Combining (\ref{eqdrhodt}) and (\ref{eqfieldr}) justifies the
	following discretization of $\partial \rho/\partial t$,
	$$
	\left (\frac{\partial \rho}{\partial t} \right )^n_{i,j}
	=-K_{i,j}^n \left [ G(R_{i,j}^n)\,\left (
	F(R_{i,j}^n)-p_{i,j}^n \right ) - (u D_1R)_{i,j}^n-(v D_2R)_{i,j}^n
	\right ]~,
	$$
	Inserting this approximation into \eqref{eq:reynolds_discrete} instead
	of $(\rho_{i,j}^n-\rho_{i,j}^{n-1})/\Delta t$ yield the One-step scheme.
	The equations are as follows:
	
	\bigskip
	
	\noindent{\bf Stage 1: Computation of $p^n$}
	\begin{multline}
	\frac{c_{i-\frac{1}{2},j}\,p^n_{i-1,j}-\left(c_{i-\frac{1}{2},j}+c_{i+\frac{1}{2},j}\right)p^n_{i,j}+c_{i+\frac{1}{2},j}\,p^n_{i+1,j}}{\Delta x_1^2} +\\ 
	+\frac{c_{i,j-\frac{1}{2}}\,p^n_{i,j-1}-\left(c_{i,j-\frac{1}{2}}+c_{i,j+\frac{1}{2}}\right)p^n_{i,j}+c_{i,j+\frac{1}{2}}\,p^n_{i,j+1}}{\Delta x_2^2} -h_{i,j}^nK^n_{i,j} G\left(R^n_{i,j}\right) p_{i,j}^n=
	\\= \frac{U}{2}\left(\frac{\rho^n_{i,j}h^n_{i,j}-\rho^n_{i-1,j}h^n_{i-1,j}}{\Delta x_1}\right)+\rho^n_{i,j}\left(D_th\right)^n_{i,j}-\\- h^n_{i,j}K^{n}_{i,j}\left(G\left(R_{i,j}^n\right)F\left(R_{i,j}^n\right)-\left(u\,D_1R\right)^{n}_{i,j}-\left(v\,D_2R\right)^{n}_{i,j}\right).\label{eq:reynolds_rp_discrete}
	\end{multline}
	Notice that the matrix to be solved for $p^n$ is not the standard one
	in Reynolds-equation solvers such as (\ref{eq:reynolds_discrete}).
	There is the additional term $-h_{i,j}^nK_{i,j}^nG(R_{i,j}^n)$ in the
	diagonal elements which is key to the enhanced stability of the scheme.
	
	\bigskip
	
	\noindent{\bf Stage 2: Computation of $R^{n+1}$}
	
	This stage of the Single-step scheme coincides with that of the
	Staggered scheme. Equation (\ref{eq:rayleigh-plesset-discrete})
	is solved to obtain $R_{i,j}^{n+1}$.
	
	\bigskip
	
	A MATLAB code for the 1D Fracture Problem along the Single-step scheme can be found in Appendix \ref{sec:appendix_code1D}.
	
	\section{Numerical Results}
	\label{sec:numerical_results}
	\subsection{The Planar Fracture}
	\label{sec:the_planar_fracture}
	
	\begin{figure}[h!]
		\centering 
		\def\svgwidth{\textwidth}	
		\includegraphics[scale=0.85]{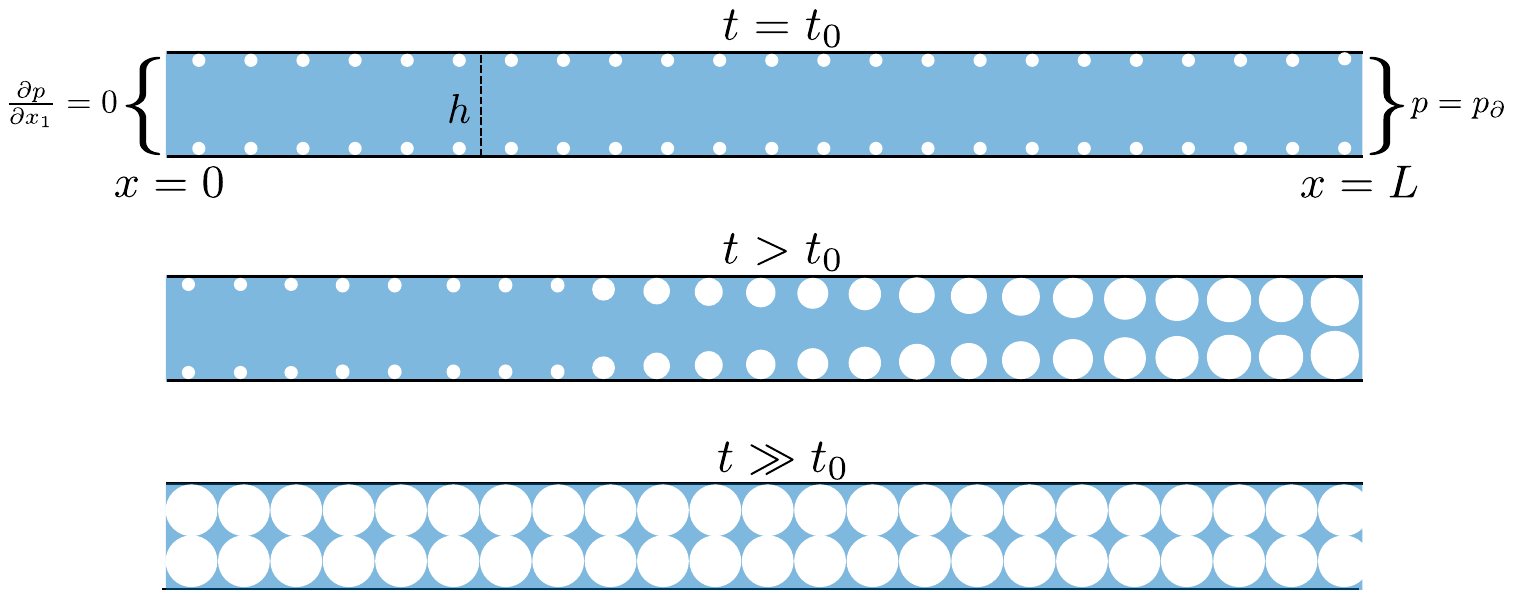}
		\caption{Scheme of the 1D fracture setup. The liquid is trapped between two parallel plates, at the left boundary a no-flux condition is imposed, at the right boundary a Dirichlet condition on pressure is imposed. The white regions represent the presence of gas-bubbles in the liquid.}\label{fig:fracture-scheme}
	\end{figure}
	
	Consider a fluid trapped between two smooth planar plates set parallel to the $x_1$-$x_2$ plane and in close proximity at distance $h$ (see Fig. \ref{fig:fracture-scheme}). The plates are infinite along the $x_2$-axis, so that the liquid pressure can be modeled by the 1D compressible Reynolds equation, which reads
	\begin{equation}
	h^2\parder{}{x_1}\left(\frac{\rho }{12\mu }\parder{p}{x_1} \right)=\parder{\rho }{t}.\label{eq:sec-fracture:reynolds-compressible-1d}
	\end{equation}
	with the boundary conditions
	$$\parder{p}{x_1}(x_1=0,t)=0,~p(L,t)=p_\partial(t).$$
	For this application it is assumed that the bubbles are attached to the surfaces ($\vec{V}=0$), and they are uniformly distributed. Denoting by $n^s_b$ the number of bubbles per unit area, the number of bubbles per unit volume is computed as $n_b=n_b^s/h$. Thus, from Eq. \eqref{eq:alpha2} we have
	\begin{equation}
	\alpha=\left(\frac{n_b^s}h\right) \frac{4}{3} \pi R^3.
	\label{eq:alpha-fracture1}
	\end{equation}
	
	If there is no presence of bubbles in the liquid, i.e., $n_b^s=0$ in  (\ref{eq:rho_alpha}) and (\ref{eq:mu_alpha}),  the fluid density $\rho$ is constant in time and so Eq. (\ref{eq:sec-fracture:reynolds-compressible-1d}) implies that the pressure along the domain is constant and equal to the boundary condition at $x_1=L$ (i.e., $p(x,t)=p_\partial(t)$). That is, the pressure in the domain adjusts instantaneously to the boundary value. 
	
	On the other hand, if $n_b^s>0$, 
	the response of the system to changes in the boundary pressure 
	is much more involved.
	Consider the system with $n_b^s$ independent of $x_1$, and with
	bubbles of initial radius $R(x_1,t=0)=R_0$ and internal pressure $P_0$. 
	This system is in equilibrium with a boundary pressure 
	$p_\partial=p_e=P_0-2\sigma/R_0$ in the sense that $\partial R/\partial t=0$
	for all $x_1$. The specific problem considered here is the 
	response of the system, initially in equilibrium, 
	after the boundary pressure $p_\partial$ is
	suddenly changed from $p_e$ ($>p_{\mbox{\scriptsize{cav}}}$) 
	to some different value $p_\partial^*<\pcav$ at $t=0$
	($p_\partial(t)=p_\partial^*,~t>0$).
	
	Since $p(x,t)$ is continuous in $x$, there will exist a region where $p<\pcav$ and thus, recalling that $\pcav$ is the minimum of $F$, 
	where the right hand side of Eq. \eqref{eq:rayleigh-plesset-withouthinertia} will be strictly positive. In that region the bubbles are expected to grow until touching one another or until filling the volume between the surfaces, at which point the model looses physical meaning. This numerical example aims at showing how this fully-gas region progresses through the domain as predicted by the RRP model. To extend the model to handle fully-gas regions, we introduce an upper limitation to the definition of $\alpha\left(R\right)$ given in Eq. \eqref{eq:alpha-fracture1}, reading
	\begin{equation}
	\alpha \left(R\right)=  \min \left\{\left(\frac{n_b^s}h\right) \frac{4}{3} \pi R^3,1\right\},
	\label{eq:alpha-fracture2}
	\end{equation}
	and also turn off the right-hand side of the Rayleigh-Plesset when
	$\alpha\geq 1$, i.e.,
	\begin{equation}
	\frac{dR}{dt}=
	\begin{cases}
	G(R)\left(F(R)-p\right)&\mbox{if }\alpha\left(R\right)<1,\\
	0&\mbox{if }\alpha\left(R\right)\geq 1.
	\end{cases}
	\label{eq:rayleigh-plesset-fracture}
	\end{equation}
	Notice that this is {\em not} a good model in general situations. In particular
	it cannot model problems in which
	a fully-gas region ($\alpha=1$) could transition back into $\alpha<1$.
	
	\subsubsection{Parameters setup}
	
	\begin{table}[h!]
		\begin{center}
			\begin{tabular}{llll}
				\toprule 
				Symbol & Value & Units & Description \\ 
				\midrule
				$\rho_\ell$ & 1000 & kg/m$^3$ & Liquid density\\ 
				$\mu_\ell$ & $8.9\times 10^{-4}$ & Pa$\cdot$s & Liquid viscosity\\
				$\mu_g^r$ & $1.81\times 10^{-5}$ & Pa$\cdot$s & Gas density\\ 
				$\rho_g$ & $1$ &  kg/m$^3$ & Gas viscosity\\
				$\kappa^s$ & $7.85\times 10^{-5}$ & N$\cdot\,$s/m & Surface dilatational viscosity\\
				$\sigma$ & $7.2\times 10^{-2}$ &N/m & Liquid surface tension\\
				
				$H$ & $10$ & $\mu$m & Gap thickness\\
				$R_0$ & $0.5$ & $\mu$m & Bubbles' radii at 1 atm\\
				$n_b^s$ & $1.91\times 10^{11}$  & m$^{-2}$ & Number of bubbles per unit area\\
				$\delta p$&$ 1.5 \cdot \pcav$& Pa & Reference value for $p_\partial^*-\pcav$ \\
				& & & (notice that $\delta p < 0$)\\
				\bottomrule
			\end{tabular} 
			\caption{Default parameters.}\label{tab:parameters-fracture-fixed}
		\end{center}
	\end{table}
	
	For the simulations presented in this Section, a set of parameters are fixed and their values are shown in Table \ref{tab:parameters-fracture-fixed}.
	The remaining free parameters, such as the length of the domain ($L$), the boundary condition ($p_\partial^*$), the bubbles internal pressure ($P_0$) and the fluid viscosity ($\mu_\ell$), are varied over wide ranges to explore the ability of the numerical methods to yield convergent solutions. We define the reference values $\delta p = 1.5 \cdot \pcav <0$ (which depends on $P_0$) and $P_0^r$ such that $P_0^r-2\sigma/R_0=1$ atm. The results shown below correspond to $\Delta x=L/1024$ and $\Delta t =1\times 10^{-6}$ s. The choose of these values is based on a mesh and time step convergence analysis such that further refinement would not be noticeable in the graphs.

	\subsubsection{Results for the Staggered scheme}
	\label{sec:results-staggered-method}
	
	The first striking result of the experiments is that if 
	$L>8.59\times 10^{-4}\,\mbox{m }$ the
	Staggered scheme produces numerical outcomes that explode exponentially
	after a few time steps. This instability cannot be avoided by
	refining the mesh or reducing the time step. 
	Furthermore, no relevant dependency of the instability on the boundary condition $p_\partial^*-\pcav$ was observed. 
	
\begin{figure}[h!]
		\centering 
		\def\svgwidth{\textwidth}	
		\includegraphics[scale=1.2]{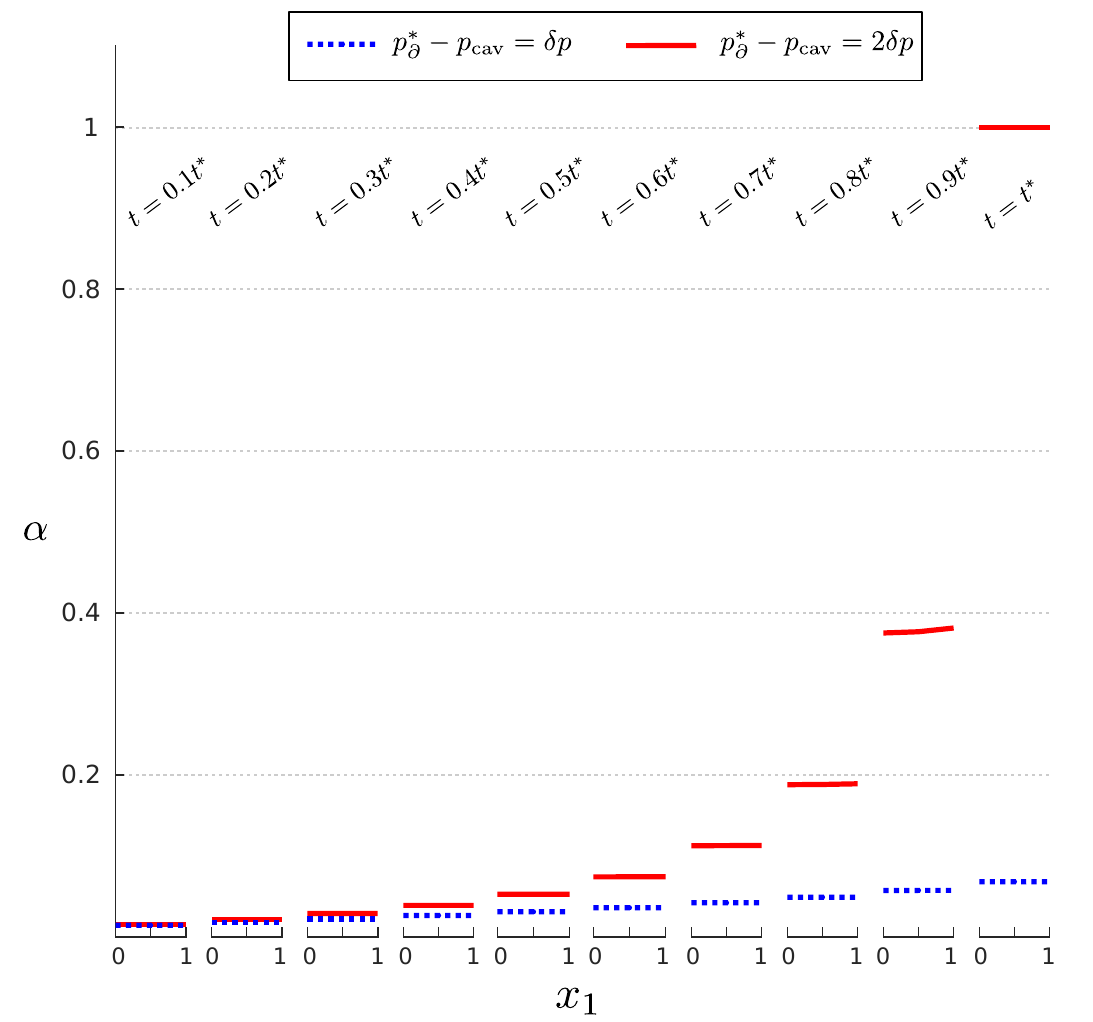}
		\caption{Results of the Staggered scheme for $L=2.15\times 10^{-4}$ m.
			Shown are some snapshots of the gas fraction profiles $\alpha(x_1,t)$, for $p_\partial^*-\pcav=\delta p,2\delta p$. The rest of the parameters were set to their default values.}\label{fig:fracture_stag_profiles_time}
	\end{figure}

	The next section explains this issue by linearizing the RRP equations and performing a zero-stability analysis \cite{leveque2007} of the Staggered scheme.
	Before that, let us provide a sample of the
	results that {\em could} be obtained to illustrate the behavior of the
	system for $L$ small. Selecting
	$L=2.15\times 10^{-4}$ m, Fig. \ref{fig:fracture_stag_profiles_time} depicts the profiles of $\alpha$ over the
	domain at several times, with two boundary conditions, $p_\partial^*=\pcav +
	\delta p$ and $p_\partial^*=\pcav + 2 \delta p$ (remember that $\delta p<0$). 
	It can be observed that for such small value of $L$ the pressure field is almost independent of $x_1$. One can also notice that the bigger the magnitude of $p_\partial^*-\pcav$, the faster the growth of $\alpha$ in time.
	
	\subsubsection{Linearizing the coupling Reynolds-Rayleigh-Plesset}
	
	Linearizing Eq. (\ref{eq:rayleigh-plesset-withouthinertia}) around an equilibrium state $(R_e,p_e)$ one gets 
	\begin{equation}
	\frac{dR}{dt}=\beta_0\left(\beta_1 R-p\right),\label{eq:rayleigh-plesset-linearized-fracture}
	\end{equation}
	where 
	$$\beta_0=\frac{R_e}{4\mu_\ell+4\kappa^s/R_e},\qquad \beta_1=F'(R=R_e).$$
	
	Please notice that $\beta_1<0$ since it is assumed that $(R_e,p_e)$ is an equilibrium state (stable branch in Fig. \ref{fig:sec:qualitative:rayleigh-plesset-p-vs-R}).
	Recalling now that $\rho/\mu$ is a function of $R$ and assuming that
	$R(x_1,t)\simeq R_e$ (as a $C^1$ function of $x_1$), one can choose
	the perturbation small enough so that
	$$\left|\parder{(\rho/\mu)}{R}\left(|R-R_e|+\parder{\left(R-R_e\right)}{x_1}\parder{p}{x_1} \right)\right|\ll \left|\frac{\partial^2 p}{\partial x_1^2}\right|~,$$
	which justifies the approximation 
	$$\parder{}{x_1}\left(\frac{\rho}{\mu}\parder{p}{x_1}\right)\approx \frac{\rho(R_e)}{\mu(R_e)}\frac{\partial^2 p}{\partial x_1^2}.$$
	Using this approximation, one can linearize Eq. \eqref{eq:sec-fracture:reynolds-compressible-1d} and use Eq. \eqref{eq:rayleigh-plesset-linearized-fracture} to get
	\begin{equation}
	\beta_2 \,\frac{\partial^2 p}{\partial x_1^2}=\frac{dR}{dt}\label{eq:reynolds-fracture-linearized}
	\end{equation}
	where $\beta_2=-\frac{h^2}{(\rho_\ell-\rho_g)\left.\parder{\alpha}{R}\right|_{R=R_e}}\frac{\rho(R_e)}{\mu(R_e)}$.
	
	Let us now discretize
	Eqs. (\ref{eq:rayleigh-plesset-linearized-fracture}) and 
	(\ref{eq:reynolds-fracture-linearized}) in space to get, respectively,
	\begin{align}
	\frac{1}{\beta_0}\frac{d \mathbf{R}}{dt}&=\beta_1\mathbf{R}-\mathbf{p}\label{eq:appendix-rayleigh-plesset-discretized}\quad\mbox{and}
	\\ \beta_2~ \mathcal{L}_{\mbox{\tiny$\Delta$}}\mathbf{p}&=\frac{d\mathbf{R}}{dt}\label{eq:appendix-reynolds-discretized},
	\end{align}
	where $\mathbf{R}(t)$ and $\mathbf{p}(t)$ are radii and pressure vectors in $\mathbb{R}^n$ and $\mathcal{L}_{\mbox{\tiny$\Delta$}}$ is the discrete Laplacian operator corresponding to the grid size $\Delta=\Delta {x_1}$. Let $\{\mathbf{g}_i\}_{i=1}^n$ be an orthonormal basis of  $\mathbb{R}^n$ formed by the eigenvectors of $\mathcal{L}_{\mbox{\tiny$\Delta$}}$ s.t.  $\mathcal{L}_{\mbox{\tiny$\Delta$}}\,\mathbf{g}_i=\lambda_i\, \mathbf{g}_i,~i=1\ldots n$. Then, $\mathbf{p}$ can be expressed as $\mathbf{p}=\sum_{i=1}^{n}\gamma_i(t)\,\mathbf{g}_i$ and so Eq. \eqref{eq:appendix-rayleigh-plesset-discretized} implies
	\begin{equation}
	\frac{1}{\beta_0}\frac{d }{dt}\left\langle\mathbf{R},\mathbf{g}_i\right\rangle=\beta_1\left\langle\mathbf{R},\mathbf{g}_i\right\rangle-\gamma_i(t),\,\qquad i=1\ldots n, \label{eq:appendix-rayleigh-plesset-discretized-modes}
	\end{equation}
	where the fact that each $\mathbf{g}_i$ is time-independent has been used. In the same way, Eq. \eqref{eq:appendix-reynolds-discretized} implies
	
	\begin{equation}
	\beta_2\,\lambda_i\,\gamma_i(t)=\frac{d }{dt}\left\langle\mathbf{R},\mathbf{g}_i\right\rangle,\,\qquad i=1\ldots n.\label{eq:appendix-reynolds-discretized-modes}
	\end{equation}
	These last two equations imply that stability analyses can be made independently for each mode $\left\langle\mathbf{R},\mathbf{g}_i\right\rangle,\,i=1\ldots n$. To simplify notation we denote $\tilde{R}=\left\langle\mathbf{R},\mathbf{g}_i\right\rangle$, $\tilde{\lambda}=\lambda_i$ and $\tilde{\gamma}(t)=\gamma_i(t)$ for some arbi{\tiny }trary index $i$. Next, the Eqs. \eqref{eq:appendix-rayleigh-plesset-discretized-modes} and \eqref{eq:appendix-reynolds-discretized-modes} are discretized in time with a constant time step $\Delta t$, such that $t^0=0$ and $t^n=n\,\Delta t$.
	
	\subsubsection{Zero-Stability of the Staggered scheme}
	
	The time discretization of Eqs. \eqref{eq:appendix-rayleigh-plesset-discretized-modes} and \eqref{eq:appendix-reynolds-discretized-modes} by the
	Staggered scheme is
	\begin{align*}
	\frac{1}{\beta_0}\frac{\tilde{R}^{n+1}-\tilde{R}^{n}}{\Delta t}&=\beta_1\tilde{R}^{n+1}-\tilde{\gamma}^{n},\\
	\tilde{\gamma}^{n}&=\frac{1}{\beta_2 \tilde{\lambda}}\frac{\tilde{R}^{n}-\tilde{R}^{n-1}}{\Delta t}.
	\end{align*}
	Substituting $\tilde{\gamma}^n$ from the latter equation into the former
	leads to
	$$\tilde{\lambda}\left(1-\Delta t\,\beta_1\,\beta_0\right)\beta_2\,\,\tilde{R}^{n+1}+\left(\beta_0-\beta_2\,\tilde{\lambda}\right)\,\tilde{R}^{n}-\beta_0\,\tilde{R}^{n-1}=0.$$
	This methodology thus corresponds to a multistep method (e.g., \cite{leveque2007} Section 5.9). Its characteristic polynomial is given by
	$$Q(\chi)=\tilde{\lambda}\left(1-\Delta t\,\beta_1\,\beta_0\right)\beta_2\,\chi^2+\left(\beta_0-\beta_2\,\tilde{\lambda}\right)\,\chi-\beta_0~.$$
	For the multistep method to be zero-stable the roots of $Q(\chi)$ must lie
	within the unit circle of the complex plane.
	Denoting these roots by $\chi_1$ and $\chi_2$, 
	$$\chi_1=1+\frac{\beta_1\,\Delta t}{{\frac{1}{\beta_0}+\frac{1}{\beta_2\,\tilde{\lambda}}}}+\mathcal{O}\left(|\Delta t|^2\right)$$
	and
	$$\chi_2=-\frac{\beta_0}{\beta_2\,\tilde{\lambda}}-\frac{\beta_0^3\,\beta_1\,\Delta t}{\beta_2^2\,\tilde{\lambda}^2}+\mathcal{O}\left(|\Delta t|^2\right).$$
	Recalling that $\beta_0>0$, $\beta_1<0$, $\beta_2<0$ and $\tilde{\lambda}<0$, one observes that $|\chi_1|<1$ for $\Delta t$ small enough. On the other hand, if $\tilde{\lambda}<\beta_0/\beta_2\,$, then $\left|\chi_2\right|$ is always bigger than one. Therefore, as the minimum eigenvalue (in magnitude) of the Laplacian operator $ \mathcal{L}_{\mbox{\tiny$\Delta$}}$ is $\tilde{\lambda}\simeq -\pi^2/(4L^2)$, instability
	is predicted for $L > \sqrt{\pi^2\beta_2/(4\beta_0)}$. The Staggered scheme is thus not zero-stable (and thus, unconditionally unstable) for $L$ large enough.
	
	For the default values of the parameters (see Table 
	\ref{tab:parameters-fracture-fixed}) it turns out that $\beta_0= 8.0\times 10^{10}$ and $\beta_2=-1.9\times 10^{-13}$ (in SI units), then one should have
	numerical instability for $L>0.024$ m. This behavior is indeed observed 
	when a {\em small} perturbation is imposed on $p_\partial(t)$. In
	Section \ref{sec:results-staggered-method} a {\em large} perturbation
	was imposed and in fact instability was observed for much smaller values
	of $L$ ($L> 8.59\times 10^{-4}$ m). The reason for this is that the
	nonlinear behavior that corresponds to cavitation amplifies
	the instability of the Staggered scheme, which can be explained from the
	change in sign of $F'(R)$ from negative to
	positive when cavitation occurs. 
	
	\subsubsection{Zero-Stability of the Single-step scheme}
	The time discretization of Eqs. \eqref{eq:appendix-rayleigh-plesset-discretized-modes}-\eqref{eq:appendix-reynolds-discretized-modes} by the Single-step scheme consists in substituting $\gamma_i(t)$ from Eq. \eqref{eq:appendix-reynolds-discretized-modes} into  Eq. \eqref{eq:appendix-rayleigh-plesset-discretized-modes}. The resulting equation reads
	$$
	\frac{\tilde{R}^{n+1}-\tilde{R}^{n}}{\Delta t}=\left({\frac{1}{\beta_0}+\frac{1}{\beta_2\,\tilde{\lambda}}}\right)^{-1}\,\beta_1\tilde{R}^{n+1}.$$
	Thus, 
	$$\tilde{R}^{n+1}=\frac{1}{1-\Delta t\,\beta_1\frac{\tilde{\lambda}\,\beta_2+\beta_0}{\tilde{\lambda}\,\beta_2\,\beta_0}}\tilde{R}^{n}.$$
	which to be zero-stable needs the factor multiplying $\tilde{R}^{n}$ to be $\leq 1$ in magnitude. But this is {\em always} true, since we have $\beta_0>0$, $\beta_1<0$, $\beta_2<0$ and $\tilde{\lambda}<0$. The unconditional zero-stability of the proposed method for the linearized RRP model has been proved. This does not guarantee numerical stability, since nonlinear effects could deteriorate its behavior. This motivates the numerical experiments below, which show that the method is stable beyond the linear regime.
	
	\subsubsection{Results for the Single-step scheme}
	\label{sec:results-single-step-method}
	\begin{figure}[h]
		\centering 
		\def\svgwidth{\textwidth}	
		\includegraphics[scale=0.85]{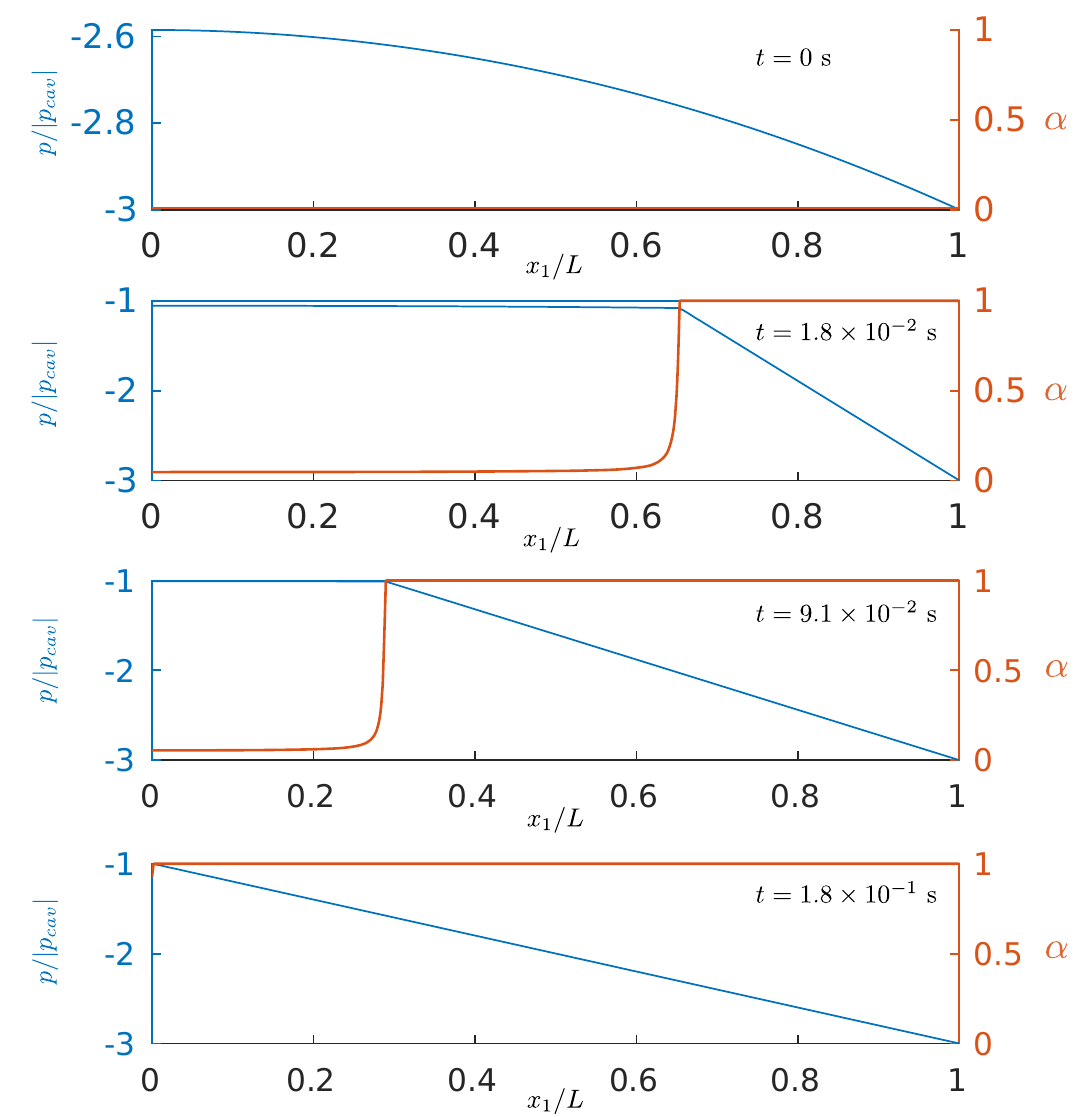}
		\caption{Gas fraction and pressure in time along the fracture setup for $L=6.9\times 10^{-3}$ m, $p_\partial^*-\pcav=4\delta p$, and the rest of the parameters set to their default values.}\label{fig:alpha_pressure_fracture}
	\end{figure}
\begin{figure}[h!]
		\centering 
		\def\svgwidth{\textwidth}	
		\includegraphics[scale=1]{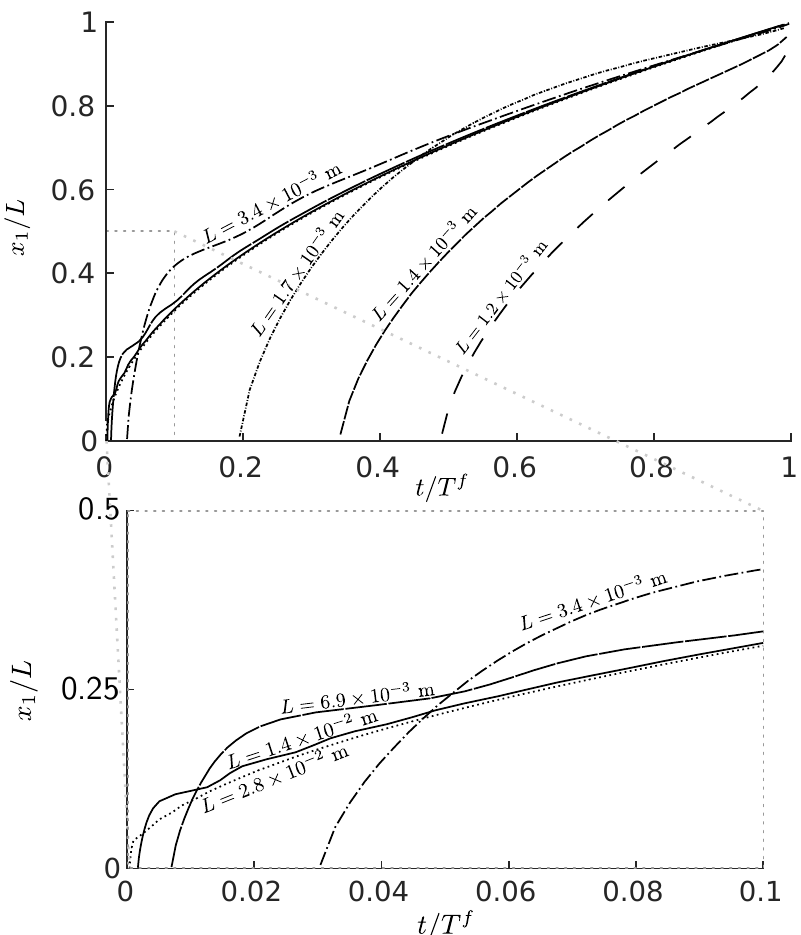}
		\caption{Non-dimensional advance of the wave for the 1D Fracture Problem for several values of $L$ and the default parameters. }\label{fig:wave_position_nondim}
	\end{figure}

	The Single-step scheme allows to perform simulations for arbitrary values of the domain length $L$.  A wave-like solution, with the cavitated region advancing towards the left, develops whenever $L>L^*\approx 1.7\times 10^{-3}$ m. An example is shown in Fig. \ref{fig:alpha_pressure_fracture} for $L=6.9\times 10^{-3}$ m.
	To depict the front advance, the position $x_1(t)$ such that $\alpha(x>x_1(t),t)=1$ and $\alpha(x<x_1(t),t)<1$ is tracked in time and the resulting curves are shown in Fig. \ref{fig:wave_position_nondim} for several values of $L$.
	Notice that 
	the time variable has been non-dimensionalized by dividing it by $T^f$, the \emph{filling time}, defined as the first time for which $\alpha=1$ on the whole domain.
	
	Interestingly, with the proposed non-dimensionalization the curves of $x_1(t)$
	converge to a unique curve when $L$ is large enough (in this case,
	for $L>L^{**}=0.0275$ m). The relative difference between the curves 
	corresponding to $L=0.0275$ m and $L=0.055$ m, for example, is less
	than 2\%. Next, a numerical study of the dependence of the filling time $T^f$ on the liquid parameters and the boundary condition $p_\partial^*$ is presented. For these analyses, the domain's length is also varied from values lower than $ L^*$ upto  values higher than $L^{**}$.
	
	\begin{figure}[h!]
		\centering 
		\def\svgwidth{\textwidth}	
		\includegraphics[scale=0.85]{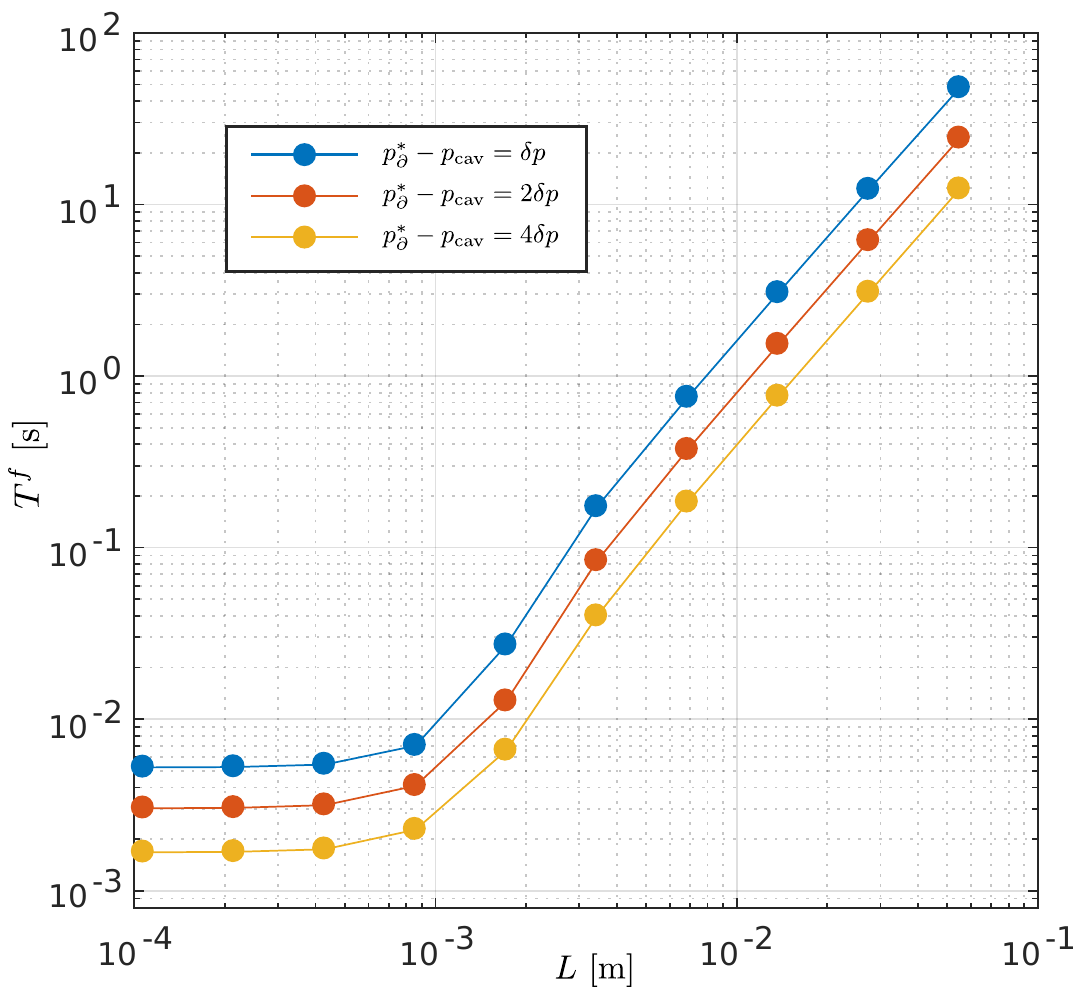}
		\caption{Filling time for several values of $L$ and $p_\partial^*-\pcav$, and the rest of the parameters set to their default values.}\label{fig:tast-varying-L-and-ppartial}
	\end{figure}
	
	Varying $p_\partial^*-\pcav$ and fixing both $P_0=P_0^r$ and $\mu_g=\mu_g^r$ the resulting filling times are shown in Fig. \ref{fig:tast-varying-L-and-ppartial} for several values of $L$. For the shorter domains, $T^f$ does not depend on the domain's length, while for the larger domains it grows quadratically with it. Notice also that $T^f$ is roughly inversely proportional to $p_\partial^*-\pcav$. 
	
\begin{figure}[h!]
		\centering 
		\def\svgwidth{\textwidth}	
		\includegraphics[scale=0.85]{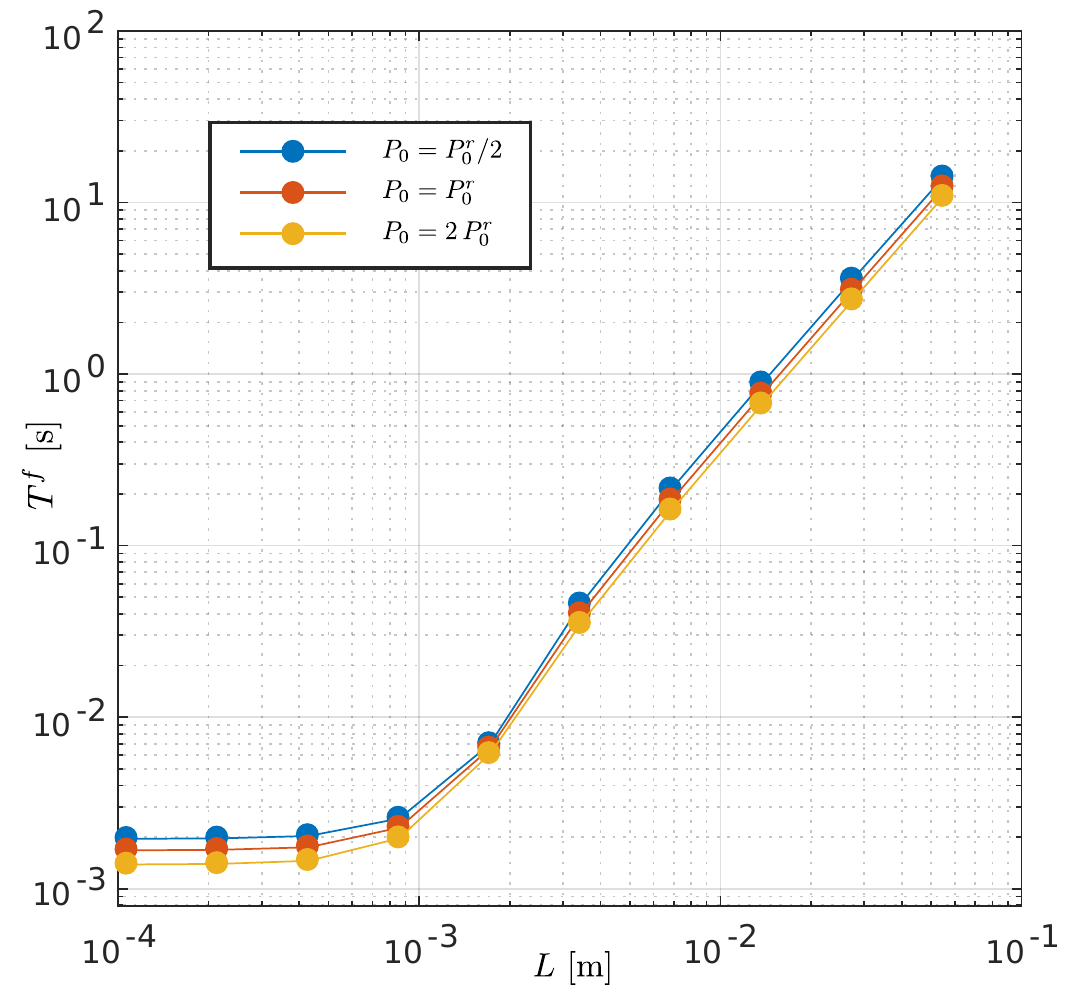}
		\caption{Filling time for several values of $L$ and $P_0$; fixing $p_\partial^*=-3.83$ atm %4\delta p$ 
			and the rest of the parameters set to their default values.}\label{fig:tast-varying-M}
	\end{figure}

	Regarding the bubbles' mass, simulations where $P_0$ is varied and $p_\partial^*$ is fixed to $-3.83$ atm are reported in Fig. \ref{fig:tast-varying-M}. This value of $p_\partial^*$ corresponds to the boundary pressure condition for the default case $P_0=P_0^r$. It is found that the filling time diminishes when augmenting $P_0$, which is expected since $\pcav$ increases monotonically with $P_0$.
	
\begin{figure}[h!]
		\centering 
		\def\svgwidth{\textwidth}	
		\includegraphics[scale=0.85]{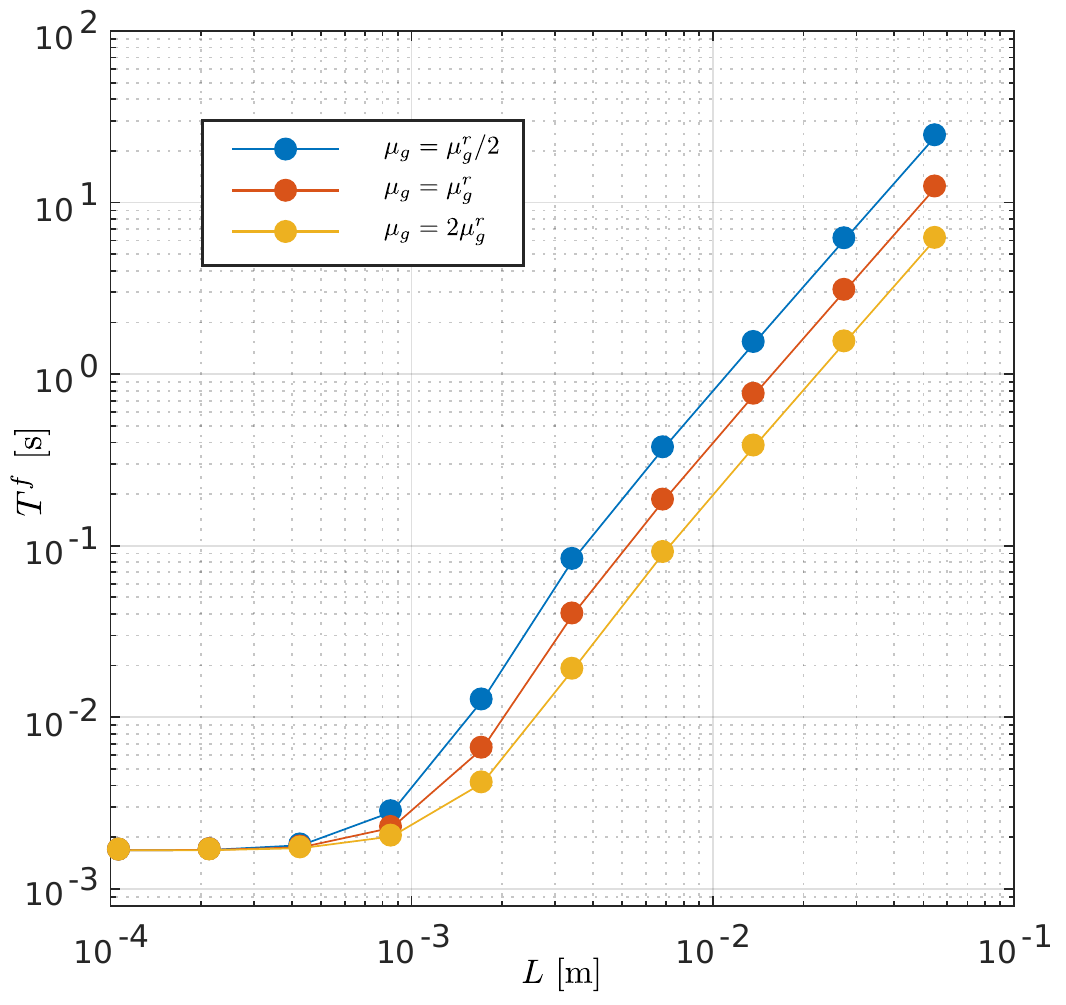}
		\caption{Filling time for several values of $L$ and $\mu_g$; fixing $p_\partial^*-\pcav=4\delta p$ and the rest of the parameters set to their default values.}\label{fig:tast-varying-muG}
	\end{figure}

\begin{figure}[h!]
		\centering 
		\def\svgwidth{\textwidth}	
		\includegraphics[scale=0.85]{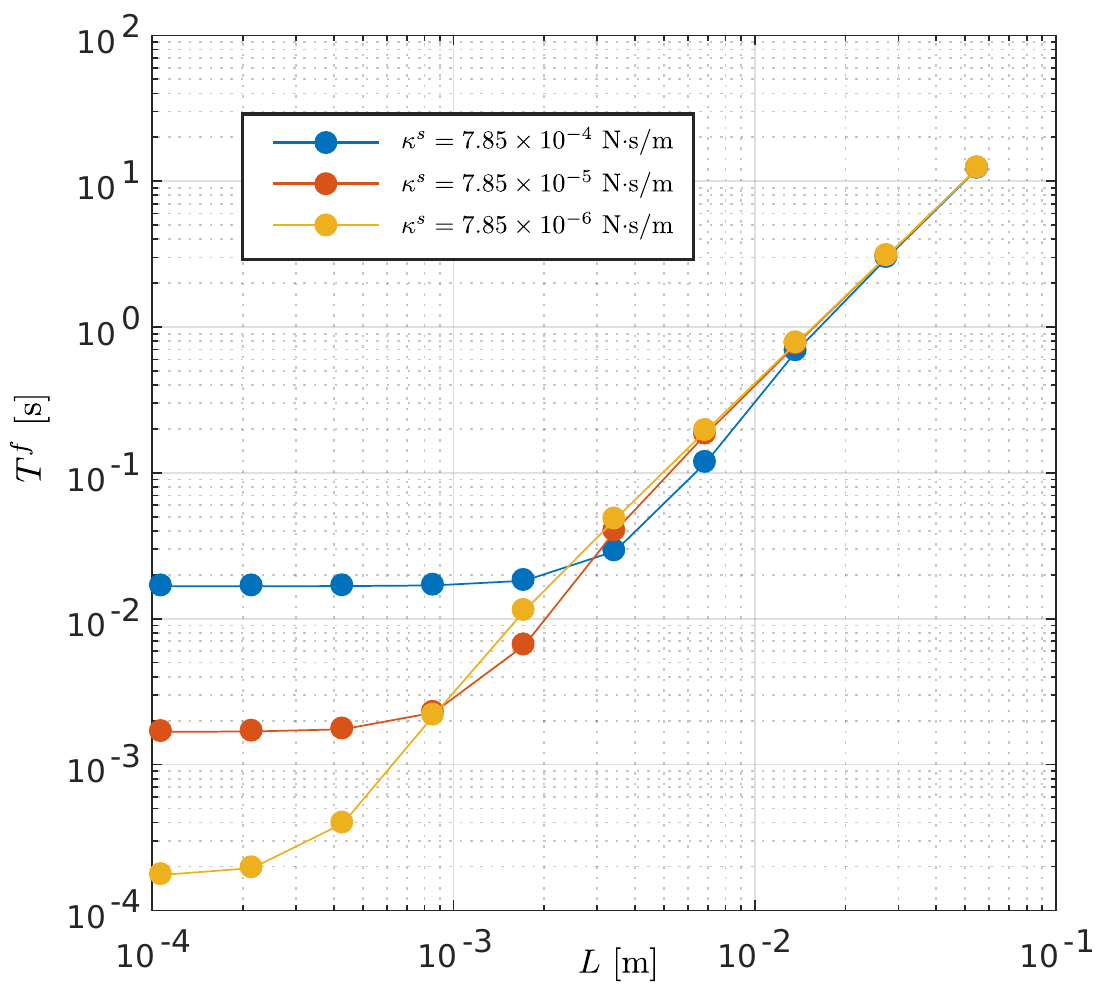}
		\caption{Filling time for several values of $L$ and $\kappa^s$; fixing $p_\partial^*-\pcav=4\delta p$ and the rest of the parameters set to their default values.}\label{fig:tast-varying-kappas}
	\end{figure}

	In the cavitated region (where $\alpha=1$) the Poiseuille flux is inversely proportional to the gas kinematic viscosity ($\mu_g/\rho_g$). This affects $T_f$ when $L$ is large enough, as shown in Fig. \ref{fig:tast-varying-muG}. Finally, the value of $\kappa^s$ is varied and the results for $T^f$ are shown in Fig. \ref{fig:tast-varying-kappas}. Notably, $T^f$ is proportional to $\kappa^s$ for small values of $L$, and independent of $\kappa^s$ for the larger domains considered. %As in the case when varying the gas kinematic viscosity,  this can be understood from the fact that the transient regime that goes from $t=0$ up-to the time when the wavefront of cavitation is totally developed is bigger (relative to $T^f$) for small domains.
	
	\subsubsection{A 2D example of the Fracture Problem}
	
	To assess the robustness of the Single-step scheme, 2D simulations of the Fracture Problem are here reported. The domain corresponds to the rectangle $[0,L]\times [0,W]$ with $L=1.25\times 10^{-2}$ m and $W=1\times 10^{-2}$ m. The grid length along $x_1$ was set to $\Delta x_1 = L/348$ and along $x_2$ to $\Delta x_2=W/256$, while the time step was fixed to $\Delta t=1\times 10^{-5}$ s. The Dirichlet condition $p_\partial^* = -2$ atm is set at $x_2=0$,
	and the null-flux condition is set at $x_2=W$, $x_1=0$ and $x_1=L$.
	
	\begin{figure}[h]
		\centering 
		\def\svgwidth{\textwidth}	
		\includegraphics[scale=0.6]{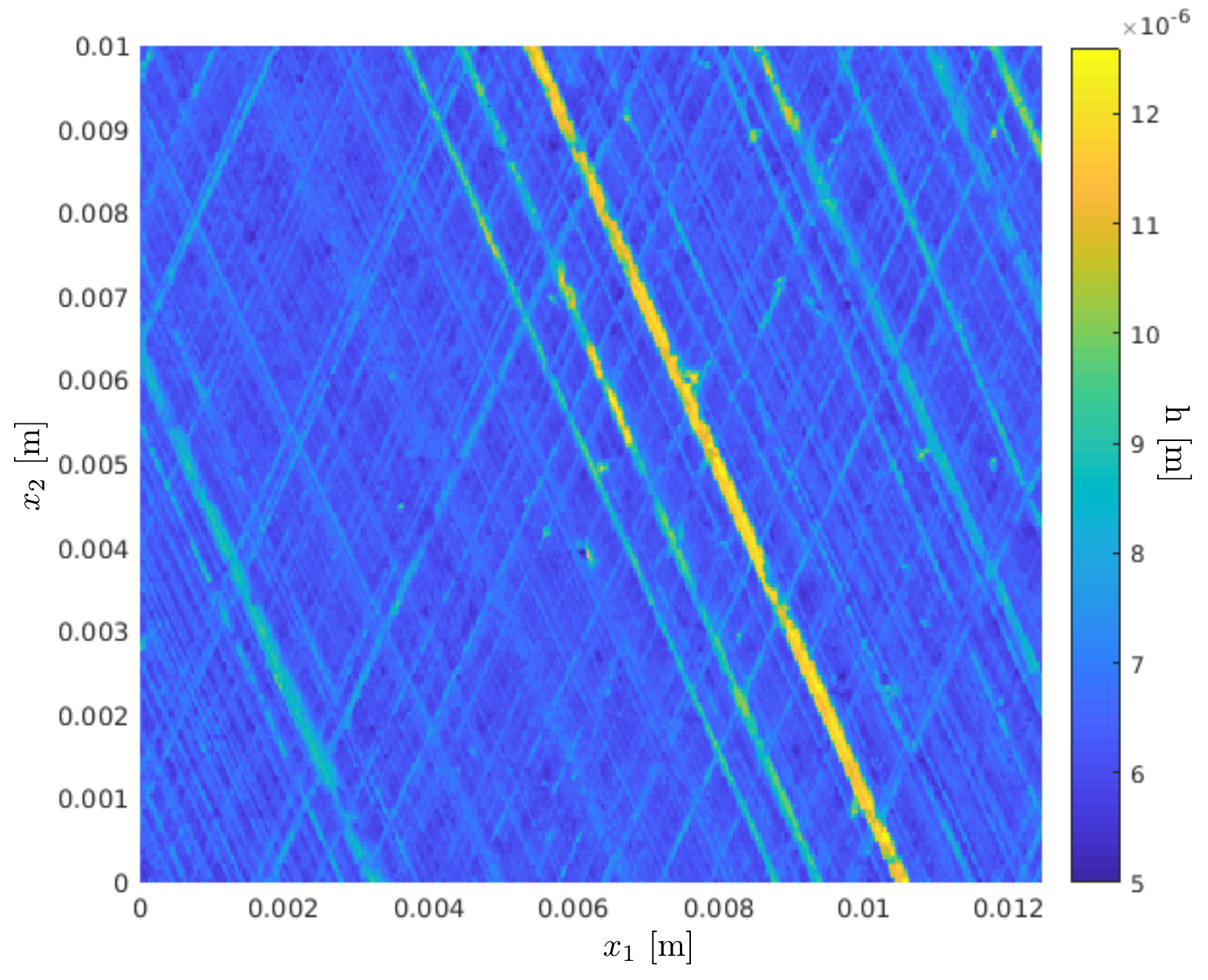}
		\caption{Height function $h(x_1,x_2)$ for the 2D Fracture Problem example.
			This gap represents the realistic distance between a flat surface and a rough surface with the presence of honed \emph{channels}.}\label{fig:fracture-2d-height}
	\end{figure}
	
	%In the previous 1D cases the gap $h$ was constant, thus it was possible to define the gas-fraction as the product of a constant times a function of $R$, i.e., $\alpha(x,t)=\alpha_0(R(x,t)/R_0)^3$, where $\alpha_0=4/3\pi R_0^3\,n_b^s/h$. 
	For the 2D cases the gap $h$ depends on $x$ as shown in Fig. \ref{fig:fracture-2d-height}.  To fix an initial gas fraction $\alpha_0=0.01$, the initial bubbles' radii are taken as $R_0=\left[\alpha_03h(x)/(4\pi n_b^s)\right]^{1/3}$. This way, $R_0$ assumes values between $0.40$ $\mu$m and $0.54$ $\mu$m (for the parameters here considered). Since it is assumed that the bubbles are in equilibrium at $1$ atm, the initial internal pressure is set according to Eq. \eqref{eq:equilibrium} or, equivalently, $P_0(R_0)=1\mbox{ atm }+2\sigma/R_0$. Thus, the cavitation pressure depends on $x$ since it varies with $P_0$ (see Eqs. \eqref{eq:Rc}-\eqref{eq:pcav}).
	
\begin{figure}[h]
		\centering 
		\def\svgwidth{\textwidth}	
		\includegraphics[scale=0.8]{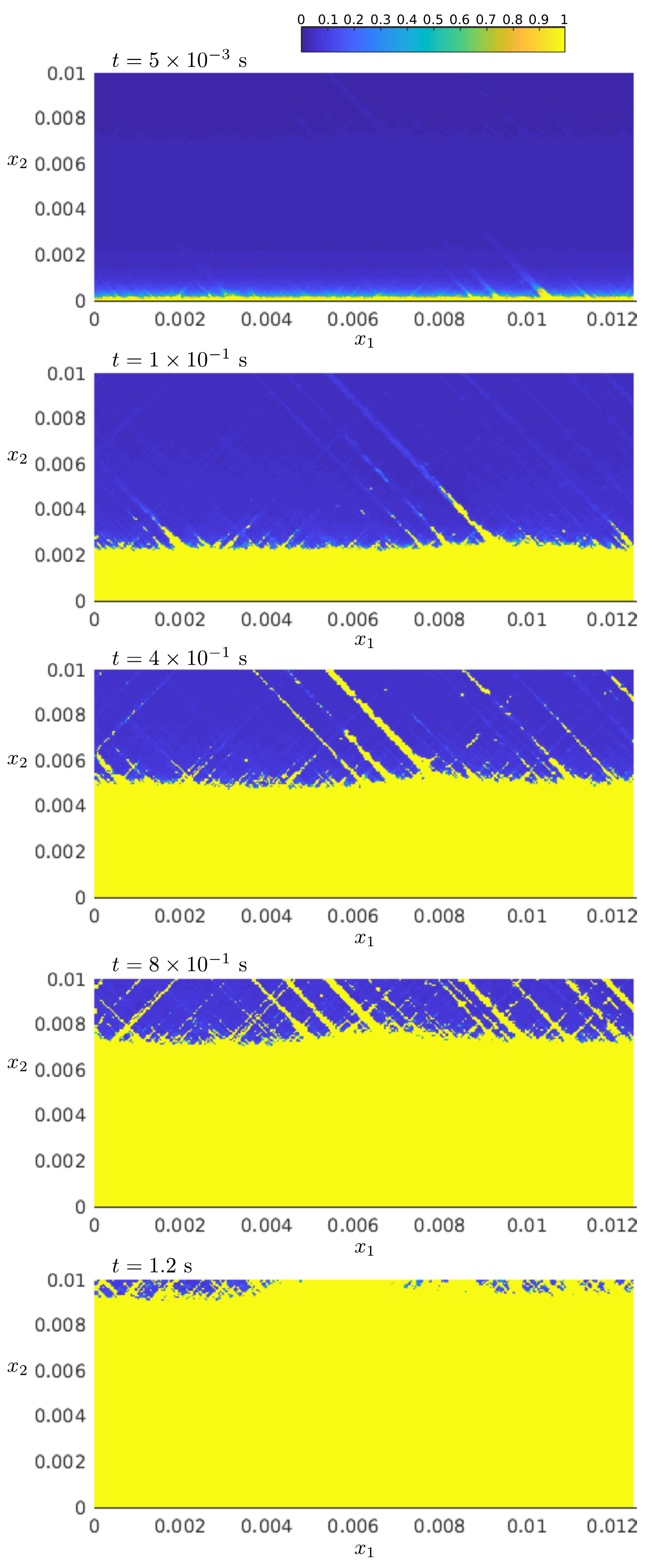}
		\caption{Evolution of the field $\alpha$ in a 2D fracture problem with a uniform initial gas fraction $\alpha_0$.}\label{fig:fracture-2d-advance-varalpha}
	\end{figure}

\begin{figure}[h]
		\centering 
		\def\svgwidth{\textwidth}	
		\includegraphics[scale=0.35]{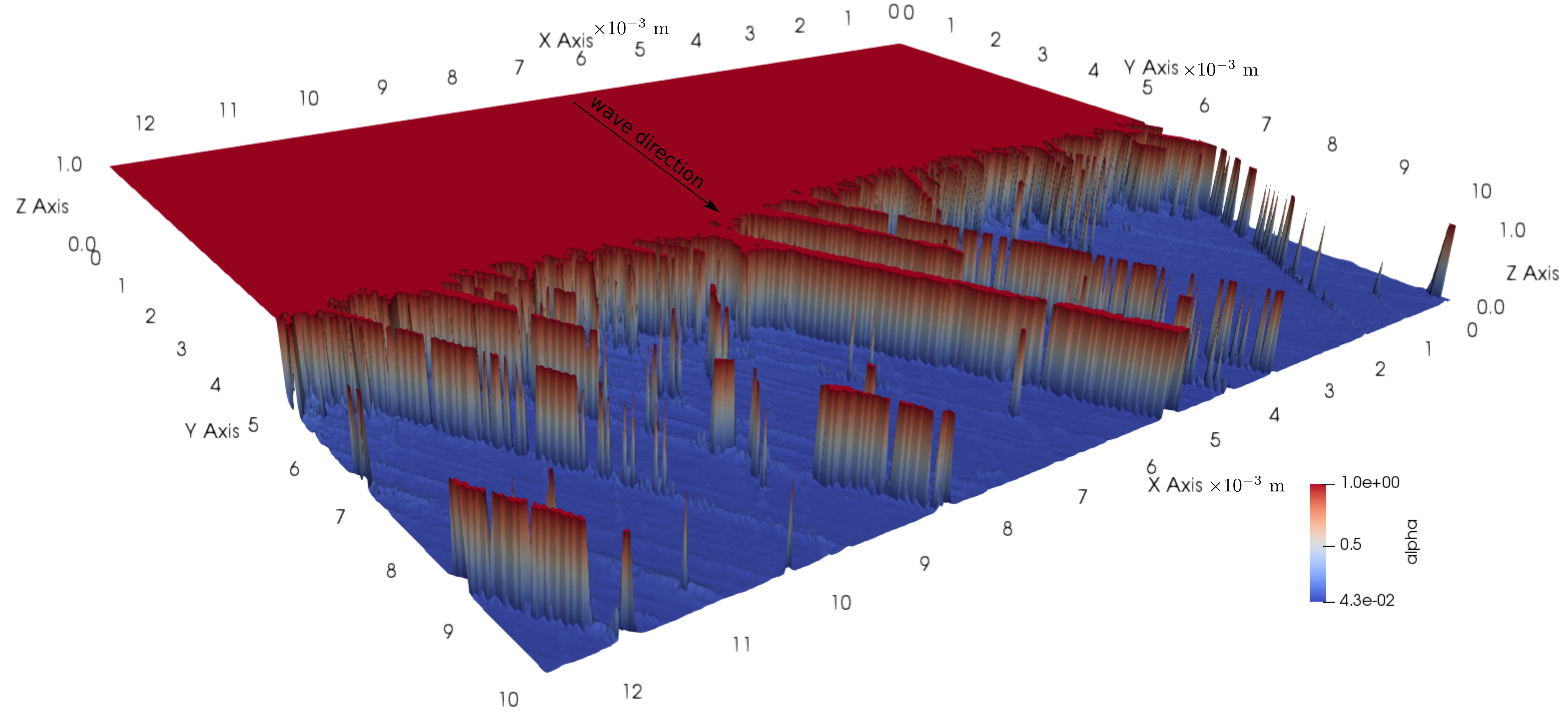}
		\caption{Three dimensional view of the gas-fraction field $\alpha$ at time $t=4\times 10^{-1}$ s for the 2D fracture problem with uniform initial gas fraction $\alpha_0=0.01$.}\label{fig:fracture-2d-3dview}
	\end{figure}

	The results here exposed are obtained with the Single-step scheme, since simulations with the Staggered scheme invariably crashed. In Fig. \ref{fig:fracture-2d-advance-varalpha} the advance of cavitation in time along the 2D domain is shown. The complexity of the field $\alpha$ can be also observed in Fig. \ref{fig:fracture-2d-3dview}. Notice the presence of cavitation in the crevices (regions with a higher value of $h$) even in places where the wave (traveling in the positive $x_2$-direction) has not arrived. This can be explained due to a higher cavitation pressure in these regions.
	
	{\em Remark:} Another possibility to perform these simulations is to set a constant $R_0$ and an initial gas fraction that depends on the position. For instance, $\alpha(x,t=0)=4/3\pi R_0\,n_b^s/h(x)$. The corresponding simulation yields solutions that are much smoother than the case presented and are not shown here since the objective is to test the Single-step method in a demanding situation.
	
	
	\subsection{The Journal Bearing}
	\label{sec:the-journal-bearing}
	
\begin{figure}[h!]
		\centering 
		\def\svgwidth{\textwidth}	
		\includegraphics[scale=0.85]{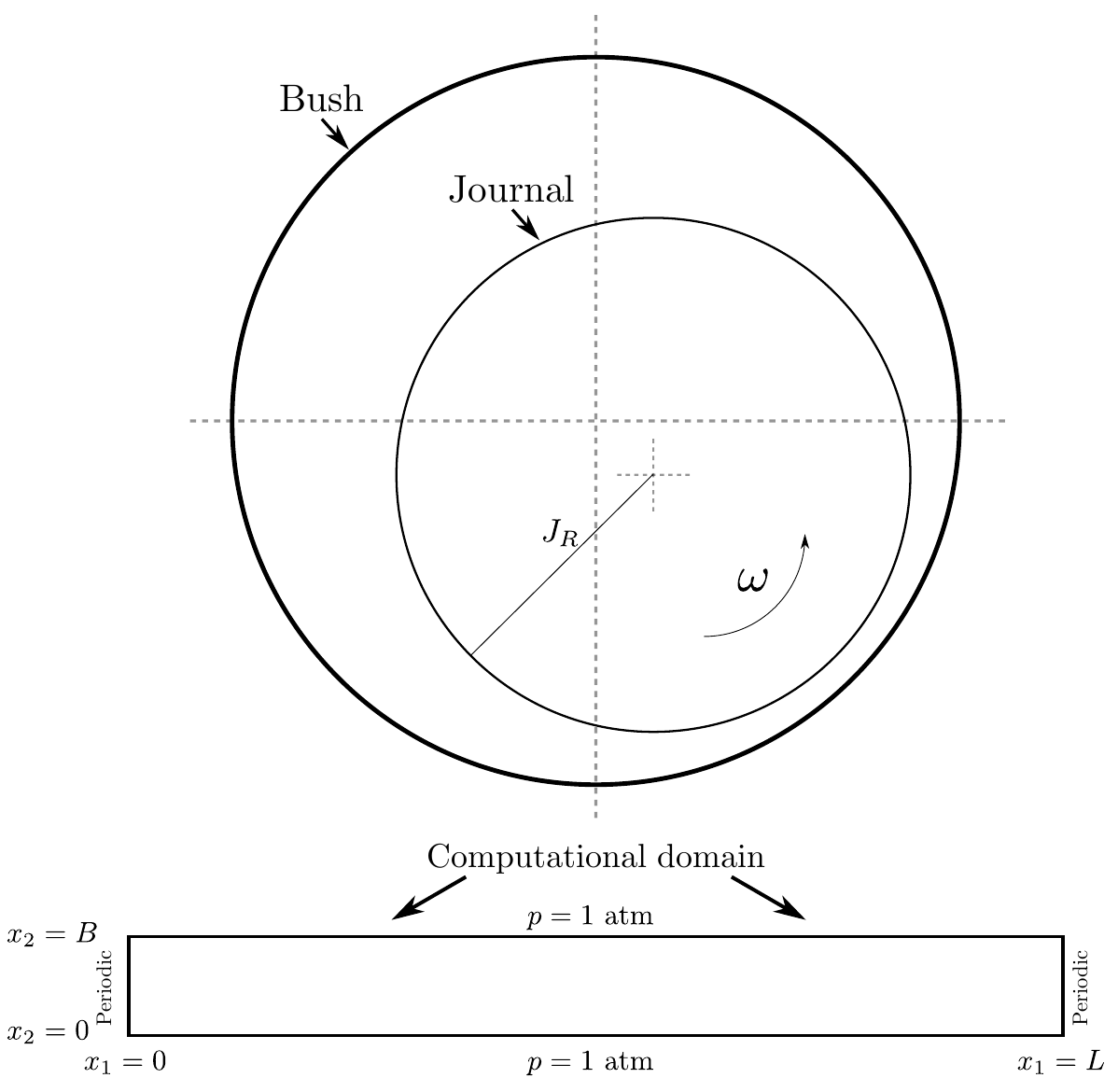}
		\caption{Scheme of the Journal Bearing.}\label{fig:journal-scheme}
	\end{figure}

	In this section results of simulations of the Journal Bearing mechanism (see Fig. \ref{fig:journal-scheme}) are presented. This problem is a typical benchmark and has already been used by other authors \cite{Natsumeda1987,Someya2003,Snyder2016,Snyder2017}. For this application, the transport of bubbles is incorporated by setting $\vec{V}=\left(\eta U,0\right)$ with $\eta\in [0,1]$, and the bubbles are assumed to be uniformly distributed in the fluid at the initial time $t=0$. The geometrical and fluid/gas parameters are shown in Table \ref{tab:parameters-journal}. As in \cite{Snyder2016}, the initial bubbles' radii is set to $R(x,t=0)=R_0=0.385$ $\mu$m. The number of bubbles per unit volume $n_b$ is assumed to be constant in space and time. Let us observe that, under this setting and by means of Eq. \eqref{eq:alpha2}, to fix $n_b$ is equivalent to fixing $\alpha_0=\alpha(R_0)$.
	
\begin{figure}[h!]
		\centering 
		\def\svgwidth{\textwidth}	
		\includegraphics[scale=0.85]{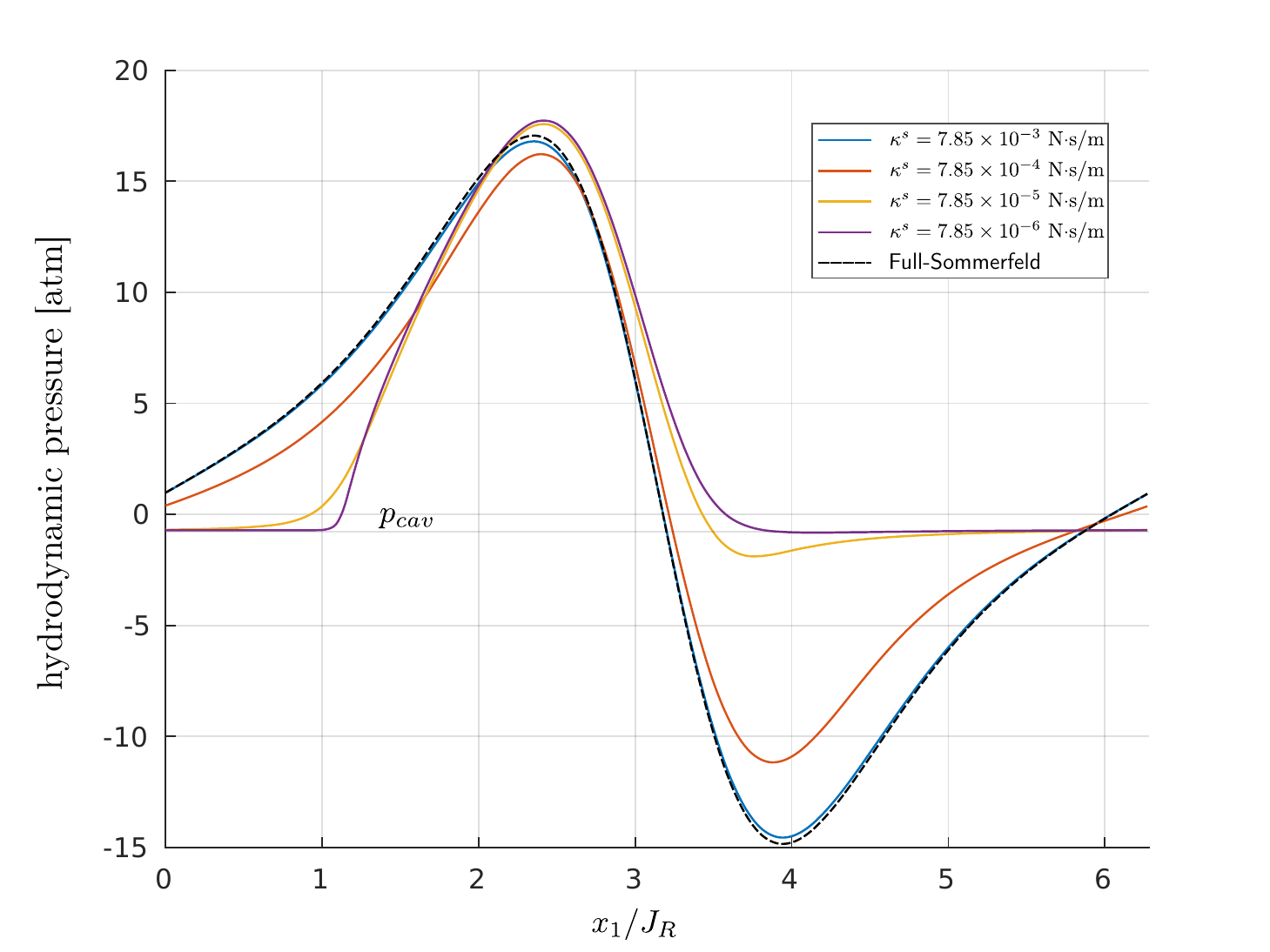}
		\caption{Pressure profiles solution of the RRP model when varying the surface dilatational viscosity $\kappa^s$ for the journal bearing rotating at 5000 rpm. Here $\alpha_0=0.05$ and $\eta=0.5$.}\label{fig:journal_p_varying_kappas}
	\end{figure}

\begin{figure}[h!]
		\centering 
		\def\svgwidth{\textwidth}	
		\includegraphics[scale=0.85]{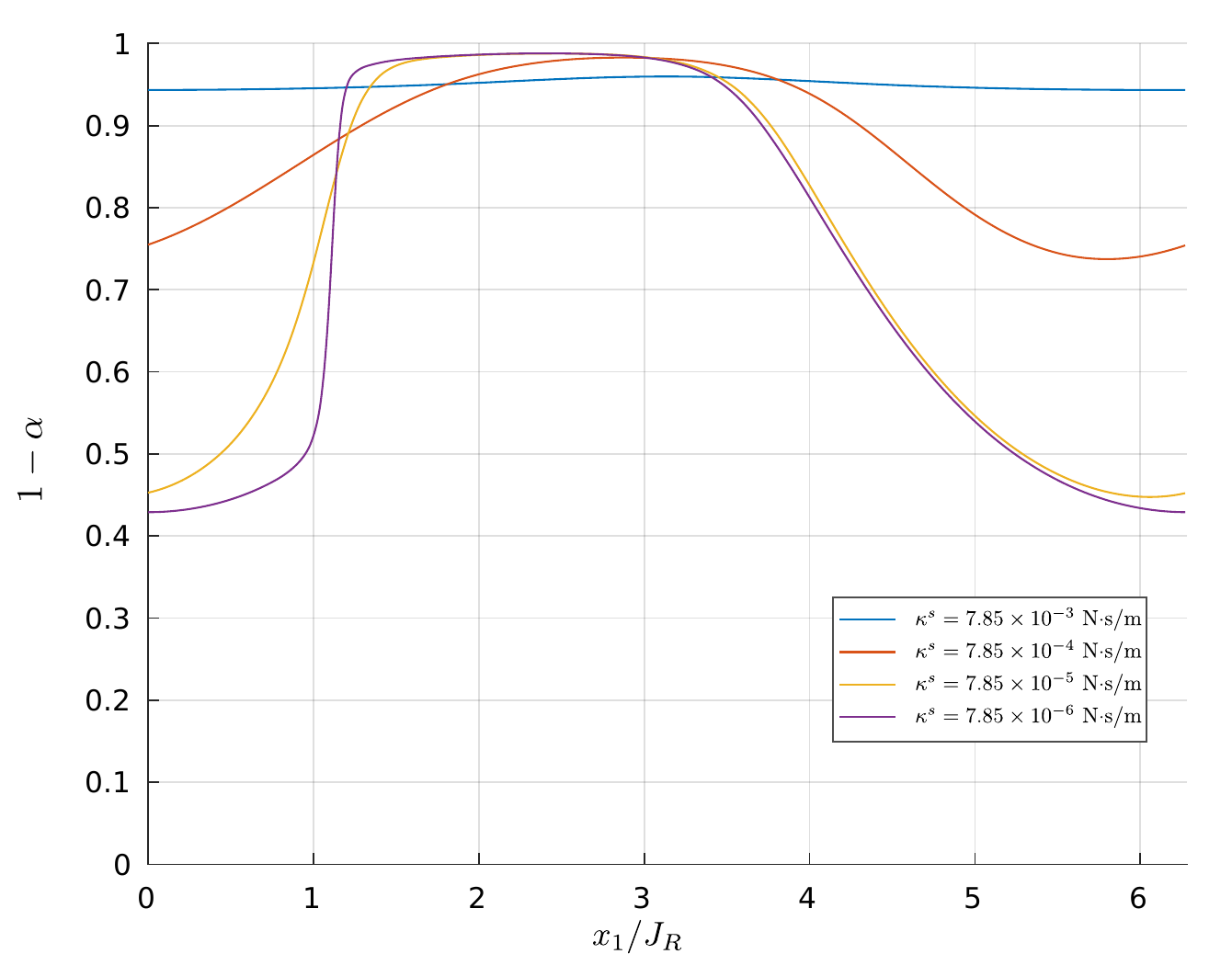}
		\caption{Liquid fraction solution of the RRP model when varying the surface dilatational viscosity $\kappa^s$ for the journal bearing rotating at 5000 rpm. Here $\alpha_0=0.05$ and $\eta=0.5$.}\label{fig:journal_theta_varying_kappas}
	\end{figure}

		While traveling through the domain, the bubbles are contracted or expanded depending on the sign of $F(R)-p$ in Eq. \eqref{eq:rayleigh-plesset-withouthinertia}. Their evolution is strongly dependent on the surface dilatational viscosity $\kappa^s$ \cite{Snyder2016}, and so are the pressure and gas fraction fields. Considering the same problem, Natsumeda and Someya set $\kappa^s$ to $7.85\times10^{-4}$ N$\cdot$s/m \cite{Natsumeda1987}. Here we explore the range $7.85\times10^{-6}$ to $7.85\times10^{-3}$ N$\cdot$s/m, 
with the results shown in Figs. \ref{fig:journal_p_varying_kappas} and \ref{fig:journal_theta_varying_kappas}. It is observed that for $\kappa^s=7.85\times10^{-3}$ N$\cdot$s/m the liquid fraction $1-\alpha$ is almost constant throughout the domain and thus the pressure profile is similar to the full-Sommerfeld curve.
For lower values of $\kappa^s$ the liquid fraction shows significant inhomogeneities which very much suggest the appearance of a {\em cavitated region} (low liquid fraction, quasi-uniform pressure). This is further discussed in 
	section \ref{sec:comparison}.

\bigskip

\noindent{\em Remark:} The results above are qualitatively similar to those reported by Snyder et al. \cite{Snyder2016} for the same journal geometry, rotational speeds, and fluid/gas physical properties. However, our results for $\kappa^s=7.85\times 10^{-6}$ N$\cdot$s/m best agree with theirs for $\kappa^s=7.85\times 10^{-4}$ N$\cdot$s/m. Similar differences on $\kappa^s$ were observed when trying to reproduce other results in their article. This difference may possibly arise from differences in
the definition of $\alpha$ in terms of $R$.

	
	\begin{table}[h!]
		\begin{center}
			\begin{tabular}{llll}
				\toprule 
				Parameter & Value & Units & Description  \\ 
				\midrule
				$\rho_\ell$ & 854 & kg/m$^3$ & Liquid density \\ 
				$\mu_\ell$ & $7.1\times 10^{-3}$ & Pa$\cdot$s & Liquid viscosity\\
				$\rho_g$ & $1$ &  kg/m$^3$ & Gas density\\
				$\mu_g$ & $1.81\times 10^{-5}$ & Pa$\cdot$s & Gas viscosity\\ 
				$\kappa^s$ & $\approx 10^{-4}$ - $10^{-6}$ &N$\cdot$s/m& Surface dilatational viscosity \\
				$\sigma$ & $3.5\times 10^{-2}$ &N/m& Liquid surface tension \\
				$p_\partial^* $ & 1 & atm & $p_\partial(t)$ for $t>0$ \\
				$p_e$ & $1$ & atm & Bubbles' equilibrium pressure \\
				$R_0$ & $0.385$ & $\mu$m& Bubbles' radii at 1 atm\\
				$\alpha_0$ & $0.05$ - $0.1$ & & Initial bubble's density\\
				$J_w$ & $25.4\times 10^{-3}$ & m & Journal width \\ 
				$J_r$ & $25.4\times 10^{-3}$ & m & Journal radius \\ 
				$J_c$ & $0.001\cdot J_r$ & m & Journal clearance \\ 
				$J_\epsilon$ & $0.4\cdot J_c$ & m & Journal eccentricity \\ 
				\bottomrule
			\end{tabular} 
			\caption{Parameter values for the Journal Bearing.}\label{tab:parameters-journal}
		\end{center}
	\end{table}
	
	\subsubsection{Stability and convergence}
	To test the stability of both methods a series of simulations were performed for $\kappa^s=7.85\times 10^{-4}$, $7.85\times 10^{-5}$ and $7.85\times 10^{-6}$ N$\cdot$s/m, rotational speeds of $1000,2000$ and $4000$ rpm, $\eta=0,0.5,1$, and $\alpha_0=0.1$ which gives a total of 27 configurations. The mesh adopted was $512\times 64$, but the same conclusions were obtained on other meshes. The time step was adjusted so that CFL$\simeq 1$.
	
	The Single-step scheme exhibits stable behavior for all of the tested configurations, reaching a stationary solution in finite time. On the other hand, the Staggered scheme fails to provide stable solutions in most of the cases. Only for $\kappa^s=7.85\times 10^{-4}$  N$\cdot$s/m and $\eta\in \{0.5,1\}$ the Staggered scheme behaves stably. The instabilities persist even if the time step is reduced a thousand times with respect to the unit-CFL value. 
	
\begin{figure}[h]
		\centering 
		\def\svgwidth{\textwidth}	
		\includegraphics[scale=0.8]{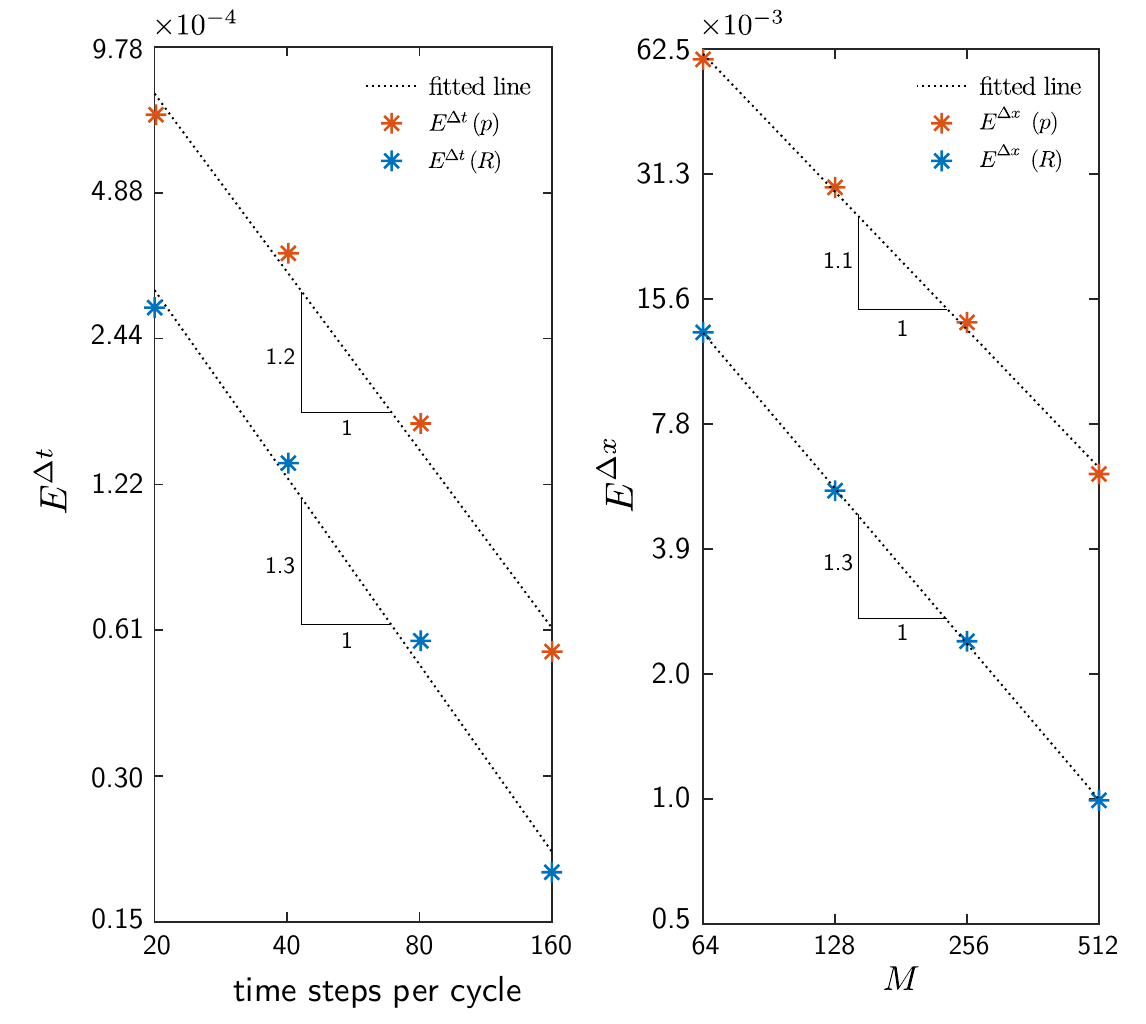}
		\caption{Convergence analysis for the journal bearing at 2000 rpm, with $\kappa^s=7.85\times 10^{-4}$ N$\cdot$s/m, $\alpha_0=0.1$ and $\eta=0.5$. Left: Time discretization error measure. Right: Space discretization error measure ($M$ is the number of grid cells along the circumferential direction).
The triangles indicate the slope of the fitted lines.}\label{fig:convergence}
	\end{figure}

	A convergence analysis is now presented for the Single-step scheme. This analysis is done for the journal rotating at 2000 rpm, with an initial gas fraction $\alpha_0=0.1$ and $\kappa^s=7.85\times 10^{-4}$ N$\cdot$s/m. To test the dependence of the solutions on the time step, the grid size along $x_1$ is set to $\Delta x_1=2\pi\,J_R/512$, and along $x_2$ to $\Delta J_w/64$. A reference solution , denoted by $(\pref,\Rref )$, is computed by setting $\Delta t$ to 640 time steps per cycle and running the simulation until $t=0.06$ s. The measure of
the temporal discretization error for a variable $f$ (which can be $p$ or $R$)
is defined as
	$$E^{\Delta t}\left(f\right)=\frac{\left\lVert f^{\Delta t}(t=0.06 \mbox{s})-\fref\right\rVert_2}{\left\lVert \fref\right\rVert _2}~,$$
where $f^{\Delta t}$ is the numerical solution computed with time step $\Delta t$.
	The results are shown on the left side of Fig. \ref{fig:convergence}, 
with strong evidence of a convergence rate of order $\approx 1$.
	
	Regarding the convergence of the discretization in space, it is
studied for the stationary solution ($t=+\infty$) to avoid interference with
time discretization errors. A sequence of nested meshes
is built by setting $\Delta x_1=2\pi\,J_R/M$ and $\Delta x_2=8J_w/M$, with
$M=64,128,256$, etc. 
The reference solutions $\pref$ and $\Rref$ are computed by setting $M=2048$.
The measure of the spatial discretization error is
	$$E^{\Delta x}\left(f\right)=\frac{\lVert f^M(t=+\infty)-\fref \rVert_2}{\left\lVert \fref \right\rVert _2}~,$$
where $f^M$ is the numerical solution computed with the grid corresponding to $M$.
The empirical convergence order as the spatial mesh is refined is of order $\approx 1$, as shown in the right side of Fig. \ref{fig:convergence}.

Up to our knowledge, this is the first numerical convergence study of
algorithms for RRP coupling. It shows that the Single-step method is
indeed stable and convergent in problems with strong nonlinear effects.
The accuracy is however limited to first order in both space and time.
	
	\subsubsection{Comparison with Elrod-Adams and Reynolds models}
	\label{sec:comparison}

\begin{figure}[h!]
		\centering 
		\def\svgwidth{\textwidth}	
		\includegraphics[scale=0.85]{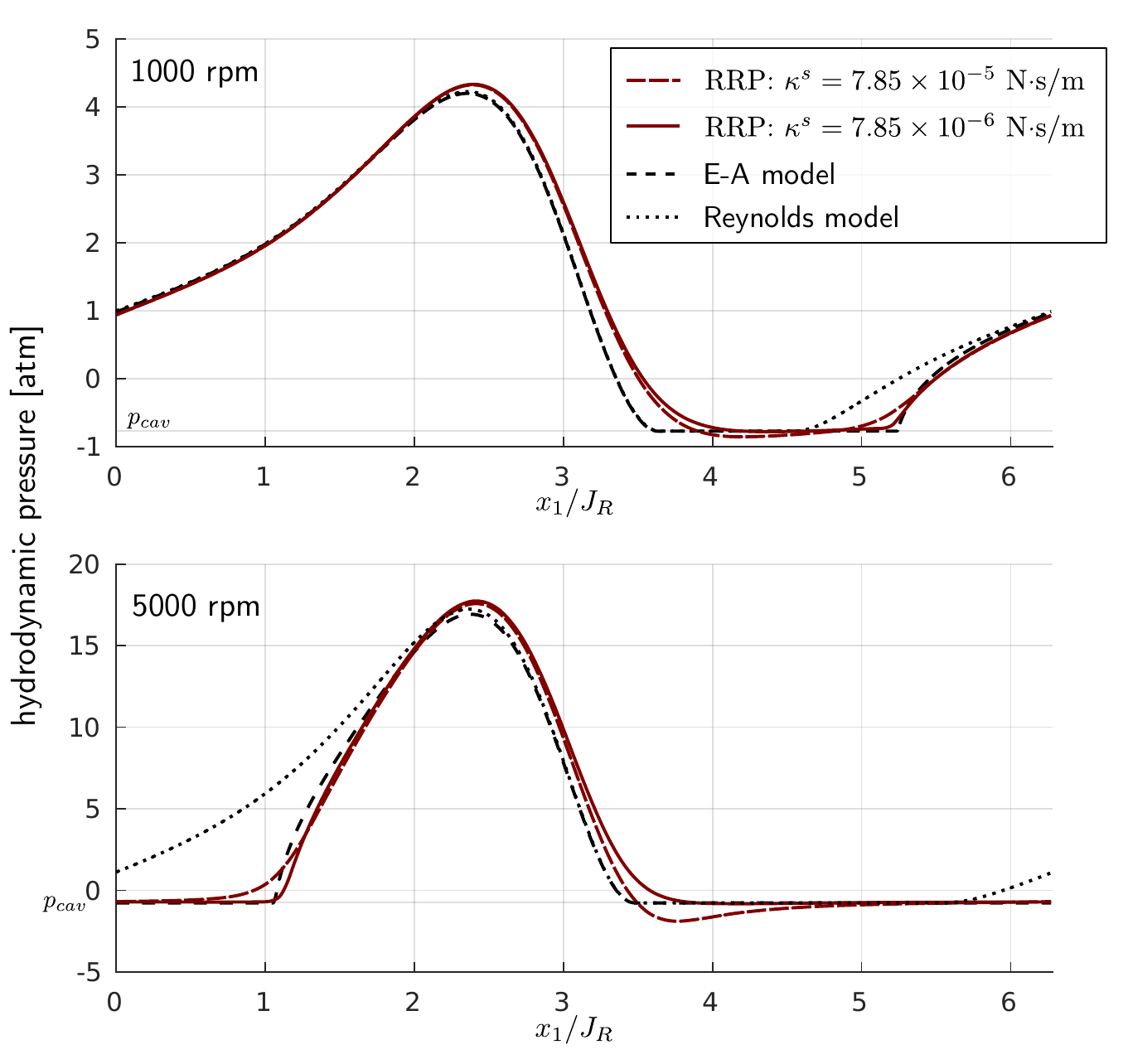}
		
		\caption{Pressure profiles obtained with the RRP, Elrod-Adams and Reynolds models for a journal bearing rotating at 1000 and 5000 rpm. Here $\alpha_0=0.05$, $\kappa^s=7.85\times 10^{-5},7.85\times 10^{-6}$ N$\cdot$s/m and $\eta=0.5$.}\label{fig:RRP-EA}
	\end{figure}

\begin{figure}[h!]
		\centering 
		\def\svgwidth{\textwidth}	
		\includegraphics[scale=0.85]{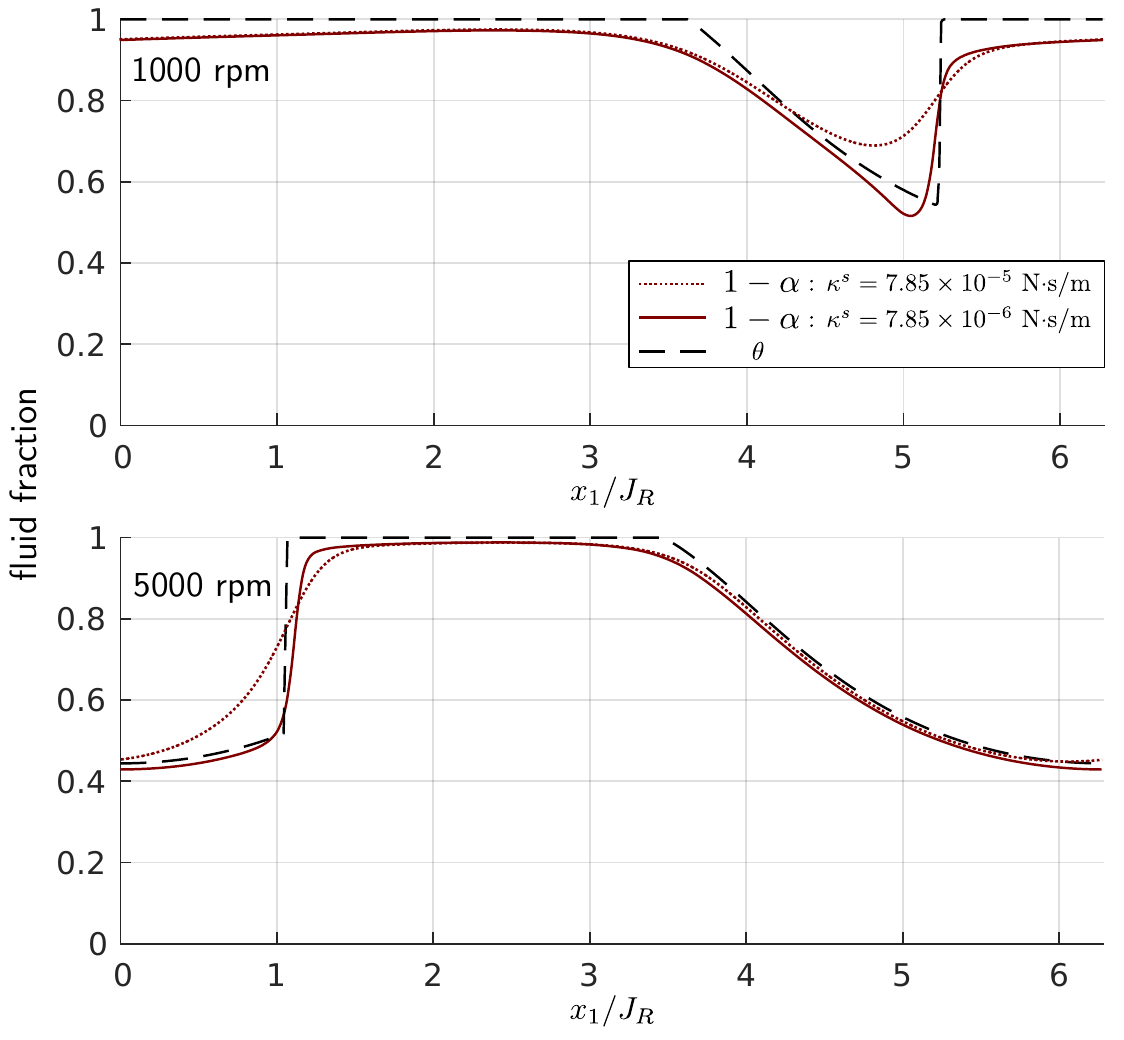}
		
		\caption{Fluid fractions obtained along the RRP coupling, Elrod-Adams and Reynolds models for a journal bearing rotating at 1000 and 5000 rpm. Here $\alpha_0=0.05$, $\kappa^s=7.85\times 10^{-5},7.85\times 10^{-6}$ N$\cdot$s/m and $\eta=0.5$.}\label{fig:thetacomparison}
	\end{figure}

	When the value of  $\kappa^s$ is small enough (e.g., $\kappa^s=7.85\times 10^{-6}$ N$\cdot$s/m) the pressure profiles that develop in the journal bearing are observed to satisfy the condition $p\geq \pcav$, with $\pcav$ computed from \eqref{eq:Rc}-\eqref{eq:pcav}. In fact, a large region where $p\simeq \pcav$ is observed,
which resembles the cavitation regions predicted by more traditional models.
This motivates to incorporate $\pcav=-0.77$ atm into the Elrod-Adams and Reynolds cavitation models in order to perform comparisons with the RRP model. Doing so, the resulting pressure profiles are shown in Fig. (\ref{fig:RRP-EA}) for rotating speeds of 1000 and 5000 rpm and $\kappa^s=7.85\times 10^{-5},7.85\times 10^{-6}$ N$\cdot$s/m. Notice that the rupture point for both the Elrod-Adams and Reynolds models are the same (which is a well-known fact), while for the RRP coupling that point is placed further along the fluid's movement direction. On the other hand, it is also known \cite{ausas07} that the Reynolds model fails to predict accurately the reformation point when compared to a mass-conserving model. Remarkably, when $\kappa^s$ is small enough the RRP model predicts a reformation point similar to that of the Elrod-Adams model. Furthermore, Fig. (\ref{fig:thetacomparison}) shows the comparison of the fluid fraction produced by the RRP model, $1-\alpha$, with the fluid fraction produced by the Elrod-Adams model, $\theta$. Qualitatively both fluid fraction fields are similar, the one corresponding to the RRP model being a regularized version of the other, in some sense. Notice that increasing $\kappa^s$ to $7.85\times 10^{-5}$ N$\cdot$s/m significantly reduces the similarities between the two models.
	
	Let us remark that the results shown in these last comparisons were obtained with a mesh having $\Delta x_1 = 2\pi J_r / 512$ and $\Delta x_2 = J_w/ 64$ (i.e., $M=512$) and with the time step is fixed to $400$ steps per cycle (CFL=1.3).
	
	\section{Conclusions} \label{sec:conclusions}
	
A stable numerical method for the RRP model (the Single-step scheme) has been proposed and compared with a strategy used in recent works (the Staggered scheme). 
A linear perturbation analysis showed that the Single-step scheme is unconditionally zero-stable, while the zero-stability of the Staggered scheme depends upon the geometrical characteristics of the mechanical system considered.
The behavior in the nonlinear range was assessed numerically, considering
two problems: A ``Fracture problem'', in which pressure build-up takes place
solely by the expansion of the bubbles (with no Couette fluxes or squeeze effects), and the well-known Journal Bearing problem.
It was found that the Single-step scheme is convergent with first order in both space and time and quite robust, allowing to perform simulations for a wide range of parameters both in the 1D and 2D settings. 
	
The simulations of the Journal Bearing also included a comparison to the Elrod-Adams model. Good agreement between both models was found when the surface dilatational viscosity is small enough. In particular, the liquid fraction $1-\alpha$ from the RRP model is quite close to the fluid fraction $\theta$ from the Elrod-Adams model. To our knowledge, this is the first time such a comparison is made and further work is under way to obtain a better insight into the relation between both models. 
	
	\section*{Acknowledgments}
	
	The authors thank the financial support of this work provided by CAPES (grant PROEX-8434433/D), FAPESP (grant 2013/07375-0) and CNPq (grant 305599/2017-8 ).
	\newpage
	\appendix
	\section{MATLAB code for the Fracture Problem 1D}
	\label{sec:appendix_code1D}
	Below a MATLAB code for the 1D Fracture Problem is presented. Notice that the functions 'RHS' and 'FOBJ' must be defined in separated files. The parameters correspond to those of Fig. \ref{fig:alpha_pressure_fracture}.
	\small
	\begin{verbatim}
	rhoL=1000;rhoG=1;muL=8.9e-4;muG=1.81e-5;
	kaps=7.85e-5;sig=7.2e-2;
	R0=0.5e-6;P0=1e5+2*sig/R0;
	L=0.0069;H=10e-6;NT=73400;%NT
	nx=512;dx=L/(nx-1);dt=2.5e-6;
	
	Rc=(3*1.4*P0*R0^(3*1.4)/(2*sig))^(1/(3*1.4-1));
	Pc=P0*(R0/Rc)^(3*1.4)-2*sig/Rc;
	
	options=optimoptions('fsolve','Jacobian',...
	'on','Display','none','TolFun',1e-6*R0,...
	'PrecondBandWidth',0,'JacobPattern',speye(nx));
	%AUXILIARY_FUNCTIONS
	global G dG F dF alpha dalpha
	alpha=@(r)min(0.01*(r/R0).^3,1);
	dalpha=@(r)3*0.01*r.^2/R0^3.*(alpha(r)<1);
	rho=@(r)(1-alpha(r))*rhoL+alpha(r)*rhoG;
	mu=@(r)(1-alpha(r))*muL+alpha(r)*muG;
	G=@(r)r.^2./(4*muL*r+4*kaps);
	dG=@(r)(2*r)./(4*muL*r+4*kaps)...
	-(4*muL*r.^2)./(4*muL*r+4*kaps).^2;
	
	F=@(r)P0*(R0./r).^(3*1.4)-2*sig./r;
	dF=@(r)2*sig./r.^2-(P0*1.4*(R0./r).^1.4)./r;
	
	R=R0+zeros(nx,1);it=1;I=1:nx-1;
	while(it<NT)%TIME_ITERATIONS
	M=H.^3.*rho(R)./mu(R);M=0.5*(M(1:nx-1)+M(2:nx));
	%SOLVE_REYNOLDS_EQUATION
	AUX=(rhoL-rhoG)*H*dalpha(R).*G(R);
	A=sparse([I I(2:end) I I nx],...
	[I I(1:end-1) [I(2:end) nx] I nx],...
	[(1/12)/dx^2*[-[0;M(1:end-1)]-M;...
	M(1:end-1);M];-AUX(1:nx-1);1],nx,nx);
	b=[-AUX(1:nx-1).*F(R(1:nx-1)); 1.1*Pc];
	P=A\b;
	%INTEGRATE_RAYLEIGH-PLESSET_EQUATION
	R=fsolve(@(r)FOBJ(r,R,P,dt),R,options);
	it=it+1;
	end
	
	function [z, dz] = RHS(r, p)
	global G dG F dF alpha
	z=(G(r).*(F(r)-p)).*(alpha(r)<1);
	dz=(dG(r).*(F(r)-p)+G(r).*dF(r))...
	.*(alpha(r)<1);
	
	function [z, dz] = FOBJ(rnp, rn, p, dt)
	[v, dv] = RHS(rnp, p);n=numel(dv);
	z = rnp - rn - dt*v;
	dz = sparse(1:n, 1:n, 1-dt*dv(:));
	\end{verbatim}

	\bibliographystyle{unsrt}
	\bibliography{lubric2017}

\end{document}